\documentclass[acmtog,nonacm]{acmart}
\acmSubmissionID{629}

\usepackage{booktabs} %
\usepackage[ruled]{algorithm2e} %

\usepackage[english]{babel}

\usepackage{graphicx}

\usepackage{hyperref}
\usepackage[]{xcolor}
\usepackage{xspace}
\usepackage{caption}
\usepackage{subcaption}
\usepackage{soul} %
\usepackage{balance}

\newcommand{\projectpage}{https://skel.is.tue.mpg.de}

\newcommand{\changed}[1]{#1}
\newcommand{\moved}[1]{}

\SetAlFnt{\small}
\SetAlCapFnt{\small}
\SetAlCapNameFnt{\small}
\SetAlCapHSkip{0pt}

\acmJournal{TOG}

\definecolor{frontcolor}{HTML}{325ea5}
\definecolor{sidecolor}{HTML}{a58b77}
\definecolor{DeltaColor}{rgb}{0.039,0.73,0.71}
\definecolor{SigmaColor}{rgb}{0.98,0.45,0.0}
\definecolor{AlphaColor}{rgb}{0,0,0.8}
\definecolor{BetaColor}{rgb}{0.8,0,0.8}
\definecolor{GammaColor}{rgb}{0.514,0.34,0.224}
\definecolor{EpsilonColor}{rgb}{0.353,0.725,0.906}
\definecolor{PurpleColor}{HTML}{bca5ea}
\definecolor{OrangeColor}{rgb}{0.914,0.541,0.0.141}
\definecolor{GreenColor}{rgb}{0.137,0.573,0.565}
\definecolor{RedColor}{rgb}{0.949,0.275, 0.224}
\definecolor{LightCyan}{rgb}{0.88,1,1}
\definecolor{Gray}{gray}{0.85}

\definecolor{NiceGreen}{rgb}{0.137,0.573,0.300}
\definecolor{NiceBlue}{rgb}{0.137,0.300,0.600}

\newcommand{\figref}[1]{Fig.~\ref{#1}}
\newcommand{\tabref}[1]{Tab.~\ref{#1}}
\newcommand{\secref}[1]{Sec.~\ref{#1}}

\newcommand{\vect}[1]{\mathbf{#1}}
\newcommand{\mat}[1]{\vect{#1}}

\newcommand{\argmin}[1]{\underset{#1}{\operatorname{arg\,min}}}

\newcommand{\osim}{\changed{BSM}\xspace}%
\newcommand{\smpl}{SMPL\xspace}
\newcommand{\skel}{SKEL\xspace} 
\newcommand{\db}{BioAMASS\xspace} 
\newcommand{\ab}{AddBiomechanics\xspace}

\newcommand{\NbBonyLdm}{57\xspace}
\newcommand{\NbSoftLdm}{48\xspace}

\newcommand{\osso}{OSSO\xspace}

\newcommand{\pose}{\boldsymbol{\theta}}     %
\newcommand{\shape}{\boldsymbol{\beta}}                 %
\newcommand{\shapespace}{\boldsymbol{B_S}}                 %

\newcommand{\smplverts}{\vect{v}^{\text{smpl}}}
\newcommand{\smpltemplate}{\mat{T}}
\newcommand{\posedirs}{B_P}
\newcommand{\smplNbdof}{72\xspace}

\newcommand{\osimexponent}{B}
\newcommand{\scales}{\vect{s}}
\newcommand{\qosim}{\vect{q}}  
\newcommand{\josim}{\vect{J}^\osimexponent}

\newcommand{\osimjoints}{\vect{J}^\osimexponent}
\newcommand{\osimNj}{N_J}  %

\newcommand{\mosimtemplate}{\vect{m_0}} %

\newcommand{\osimtemplate}{\mat{T}^\osimexponent}

\newcommand{\targetmarkers}{\vect{m}^T}
\newcommand{\markeroffset}{\boldsymbol{\delta}}

\newcommand{\osimNbdof}{46\xspace}
\newcommand{\osimNbparts}{24\xspace}

\newcommand{\qskel}{\qosim} 
\newcommand{\jskel}{\vect{J}}

\newcommand{\skelskinverts}{\vect{v}^{\text{skin}}}
\newcommand{\skelskelverts}{\vect{v}^{\text{skel}}}

\newcommand{\jointreg}{\mathcal{J}}

\newcommand{\smpltoloccol}[1]{\textcolor{NiceGreen}{#1}}
\newcommand{\loctosmplcol}[1]{\textcolor{NiceBlue}{#1}}

\newcommand{\jregressed}{\vect{J}^{reg}}
\newcommand{\jsmpl}{\vect{J}^{smpl}}
\newcommand{\jskelfit}{\vect{J}^{skel}}

\acmJournal{TOG}
\setcopyright{none}



\begin{document}

\title{From Skin to Skeleton: Towards Biomechanically Accurate 3D Digital Humans}

\author{Marilyn Keller}
\orcid{0000-0003-2611-8595}
\affiliation{%
 \institution{Max Planck Institute for Intelligent Systems}
 \streetaddress{Max Planck Ring 4}
 \city{T\"{u}bingen}
 \postcode{72076}
 \country{Germany}}
\email{marilyn.keller@tuebingen.mpg.de}
\author{Keenon Werling}
\orcid{0000-0003-3506-7769}
\affiliation{%
 \institution{Stanford University}
 \city{Stanford}
 \state{CA}
 \country{USA}}
 \email{keenon@stanford.edu}

\author{Soyong Shin}
\orcid{0000-0002-0406-7611}
\affiliation{%
 \institution{Carnegie Mellon University}
 \city{Pittsburgh}
 \state{PA}
 \country{USA}}
\email{soyongs@andrew.cmu.edu}
\author{Scott Delp}
\orcid{0000-0002-9643-7551}
\affiliation{%
 \institution{Stanford University}
 \city{Stanford}
 \state{CA}
 \country{USA}}
\email{delp@stanford.edu}

\author{Sergi Pujades}
\orcid{0000-0002-9604-7721}
\affiliation{%
 \institution{Inria centre at the University Grenoble Alpes}
 \city{Grenoble}
 \country{France}}
 \email{spujades@tuebingen.mpg.de}
\author{C.~Karen Liu}
\orcid{0000-0001-5926-0905}
\affiliation{%
 \institution{Stanford University}
 \city{Stanford}
 \state{CA}
 \country{USA}}
\email{karenliu@cs.stanford.edu}
\author{Michael J.~Black}
\orcid{0000-0001-6077-4540}
\affiliation{%
 \institution{Max Planck Institute for Intelligent Systems}
 \city{T\"{u}bingen}
 \country{Germany}}
 \email{mjb@tuebingen.mpg.de}

\renewcommand\shortauthors{Keller, M. et al}

\begin{teaserfigure}
\centering
  \includegraphics[width=1\textwidth]{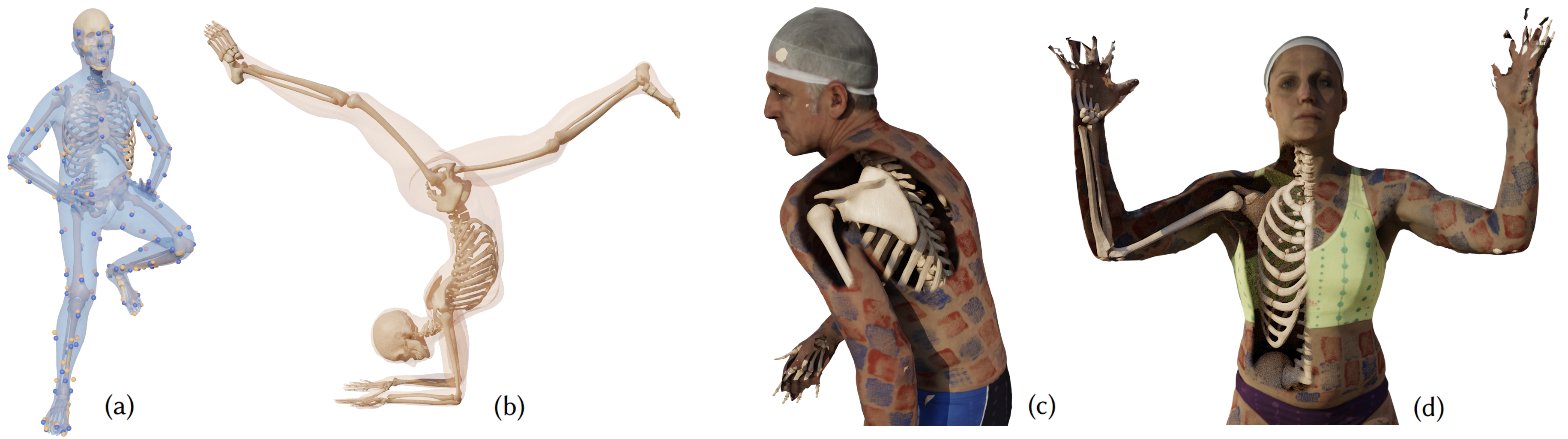}
  \caption{
  (a) We fit \changed{our new Biomechanical Skeleton Model}, \osim, to SMPL \cite{loper2015smpl} mesh sequences from AMASS \cite{mahmood2019amass}. This gives paired data enabling us to learn the  mapping from skin to skeleton. (b) We use this to create \skel, a parametric body model with skin and skeleton meshes, driven by biomechanical pose parameters and incorporating the shape space of SMPL.
  \skel is like \smpl but with more realistic degrees of freedom. 
  \changed{Fitting \skel to DFAUST scans \cite{bogo2017dfaust} results in  \skel's scapula sliding (c) and the forearms twisting appropriately (d).}
  }
  \label{fig:teaser}
\end{teaserfigure}

\begin{abstract}
Great progress has been made in estimating 3D human pose and shape from images and video by training neural networks to directly regress the parameters of parametric human models like SMPL.
However, existing body models have simplified kinematic structures that do not correspond to the true joint locations and articulations in the human skeletal system, limiting their potential use in biomechanics.
On the other hand, methods for estimating biomechanically accurate skeletal motion typically rely on complex motion capture systems and expensive optimization methods.
What is needed is a parametric 3D human model with a biomechanically accurate skeletal structure that can be easily \changed{posed}. %
To that end, we develop \skel, 
which re-rigs the SMPL body model with a biomechanics skeleton.
To enable this, we need training data of skeletons inside SMPL meshes in diverse poses.
We build such a dataset by optimizing biomechanically accurate skeletons inside SMPL meshes from AMASS sequences. 
We then learn a regressor from SMPL mesh vertices to the optimized joint locations and bone rotations.
Finally, we re-parametrize the SMPL mesh with the new kinematic parameters.
The resulting \skel model is animatable like SMPL but with fewer, and biomechanically-realistic, degrees of freedom.
We show that SKEL has more biomechanically accurate joint locations than SMPL, and the bones fit inside the body surface better than previous methods.
By fitting \skel to SMPL meshes we are able to ``upgrade" existing human pose and shape datasets to include biomechanical parameters.
\skel provides a new tool to enable biomechanics in the wild, while also providing vision and graphics researchers with a better constrained and more realistic model of human articulation.
The model, code, and data \changed{are}
available for research at \changed{\url{\projectpage}}.
\end{abstract}

\maketitle

\section{Introduction}

Human motion is captured, modeled and studied in diverse fields, including computer vision, graphics, gaming, biomechanics, medicine, ergonomics and more.
The tools and representations used, however, vary significantly. 
Vision and graphics methods often represent the articulated body pose using an approximate 3D skeleton, whereas, in biomechanics and sports medicine, an accurate kinematic skeleton is of paramount importance for disease diagnosis.
The capture methods also vary significantly.
Computer vision focuses on estimating 3D humans from images and videos while the biomechanics community focuses on highly accurate marker-based motion capture \changed{(mocap)} systems.
This paper takes a step towards combining the best of these disciplines, providing new and improved tools to each; \changed{see~\figref{fig:teaser}.}

Specifically, we focus on advances in computer vision that infer the 3D pose and shape of the human body in the form of parametric body models like SMPL \cite{loper2015smpl}.
The field has advanced rapidly and the accuracy of markerless video-based 3D motion capture is catching up with marker-based techniques. 
Unfortunately, the kinematic structure of models such as SMPL is not physically accurate, \changed{limiting applicability in biomechanics.}
On the other hand, the biomechanics field has developed detailed skeletal models \changed{to represent the anatomic motion of the knee, spine, shoulder, etc.}
The vision and graphics communities are currently not benefiting from these more accurate models of the body and its joints.

To address these issues, we unify the SMPL body model with \osim, a new \changed{Biomechanical Skeleton Model}. 
While previous work has addressed the problem of putting skeletons inside 3D body models \cite{ali2013anatomy, kelc2012zygote, shetty2023boss, keller2022osso}, such approaches have not addressed the problem of precisely locating the skeleton within a {\em moving} body.
The key challenge is the lack of training data that pairs the posed 3D human body shape with the ground-truth skeleton.
We address this by creating a novel dataset called \db. 
To create \db, we take sequences of 3D bodies from the AMASS dataset \cite{mahmood2019amass} that cover a wide range of body shapes and challenging poses.
To obtain pseudo-ground-truth skeletons, we place virtual motion capture markers on the body surface.
We then use \changed{the recent method,} %
\ab \cite{werling2022addbio}, to solve for the \changed{BSM} skeleton given the virtual markers.

With this paired dataset, we can now solve several problems that were previously impossible.
First, we train a regressor to estimate the 3D anatomical \osim joint locations of the body given a posed SMPL mesh.
Note that these locations significantly differ from the joints in SMPL.
This is useful for generating more relevant training data for 2D or 3D joint detectors,
as today, such methods are typically trained from manually labeled joints or projected SMPL joints.

Next, we re-rig the SMPL body model with the \osim biomechanical skeleton, i.e.~we use the \osim parameters to drive a SMPL mesh, and we call the resulting model \skel, which is short for ``Skeletal Kinematics Enveloped by a Learned body model''.
To do so, it is critical that the skeleton is properly scaled, located and oriented inside the SMPL body mesh.
To that end, we propose a data-driven strategy that places the bones inside the body while ensuring that their orientations are compatible with the anatomic constraints of the limbs. 
Like SMPL, \skel provides a body surface but with a skeleton inside that has biomechanical degrees of freedom.
For example, %
the spine in \skel is modeled by a spline derived from biomechanics.
Additionally, shoulders are a complicated structure that is typically crudely approximated in vision and graphics models.
\skel replaces the approximate shoulder of SMPL with a biomechanical shoulder blade \cite{seth2016scapulothoracic} that slides along an ellipsoid defined around the thorax.
The forearm rotation is another place where standard graphics models like SMPL differ from biomechanics.
Instead of a simple rotation around the elbow, \skel models the motion of the radius and ulna bones to drive forearm pronation and supination.

\skel has several uses. Specifically, we consider the problem of taking a SMPL body model and computing the correct skeleton inside.
To do so, we simply fit \skel to the posed SMPL mesh by optimizing the \skel pose to minimize the vertex-to-vertex distance between the meshes.
We apply this process to archival datasets such as 3DPW \cite{von2018recovering} and BEDLAM \cite{bedlam}. 
This effectively {\em upgrades} existing computer vision datasets to contain biomechanical ground truth, extending their use to biomechanics.
For example, one could evaluate, or learn to directly regress biomechanical parameters from video.

\changed{We evaluate two methods for estimating the skeleton from SMPL: direct regression of \osim joints from SMPL and fitting SKEL to SMPL. Accuracy is defined in terms of 3D joint location error.}
Since there is no ground-truth for this task, we take the joint locations estimated by \ab as pseudo-ground-truth.
We find that both of our methods produce significantly more accurate joint predictions than \smpl. 
We also provide extensive qualitative experiments that show the articulated structure of \skel and its use in upgrading existing human motion datasets to support biomechanics.

\skel can also be used in the other direction. Given an input skeleton mesh obtained after fitting a biomechanical model to mocap data, \skel can be used to add a plausible skin surface; this is useful for visualization of mocap data.
Since there are an infinite number of body shapes that are consistent with a given skeleton, the predicted shape can be constrained, e.g.~with the subject's weight.

To the best of our knowledge, \changed{ \skel is the first model where the body surface and anatomical skeleton are directly controlled by the same set of shape and pose parameters ($\shape, \qosim$). }
The \db dataset,
\changed{the code to create it from the AMASS dataset,}
as well as the \skel model, \changed{are} available for research purposes
at \changed{\url{\projectpage}}.

\section{Related Work}

The accurate representation and animation of human bodies play an important role in computer graphics, vision, and biomechanics. 
There have been significant recent advances in the creation of statistical body surface models, biomechanical anatomical models, and techniques for extracting these models from motion capture data. 

{\bf Body models.}
In vision and graphics, statistical body shape models are widely used \cite{allen2003space,anguelov2005scape,loper2015smpl,wang2020blsm,Pavlakos2019smplx,xu2020ghum,osman2022supr,osman2020star}.
These models are trained using 3D scans of people with many body shapes in many poses and provide an accurate representation of the human body surface.
However, their skeletal structure and their joint locations are not designed to correspond to the anatomical functional joints of the body. 
For example, their kinematic tree does not match the degrees of freedom of the human anatomic skeleton. The knee and elbow flexion, the spine, the elbow and the arm supination are typically modeled by ball joints, while those functional joints have only one major degree of freedom or are more complex than a pure rotation, such as the knee, spine, or the shoulder.

\moved{[Moved from the motion capture paragraph]}
Predicting the location of joints, such as the femur head, from 2D images of clothed individuals is \changed{inherently ill-posed because the joint location is not directly observed.}
Instead of directly estimating joints from video, one can fit, or regress, SMPL or SMPL-X body model parameters \cite{Kanazawa:CVPR:2018,Pavlakos2019smplx,Kocabas2020vibe}.
From SMPL one can then extract the 3D joint locations, but
unfortunately, SMPL joints are not anatomically correct.
In this paper, we quantify the error of the SMPL joint locations w.r.t.~a biomechanical model
and learn a regressor that better predicts the functional joint locations inside SMPL. \moved{[End of moved material]}

{\bf Biomechanical skeleton models.}
In contrast to body models, skeletal models used in biomechanics, e.g.~\citet{rajagopal2016full}, \citet{seth2016scapulothoracic}, \citet{nitschke2020efficient}, 
define the degrees of freedom of the human skeleton with a focus on anatomic realism.
This is critical for kinematic and kinetic analysis. 
The size and motion of these skeleton models are computed from optical motion capture data using optimization frameworks like OpenSim \cite{delp2007opensim} or \ab \cite{werling2022addbio}.
\changed{This is the classical approach} in biomechanics for measuring the precise location of the functional joints.

{\bf Motion capture. }
While marker-based motion capture (mocap) is the preferred method for analyzing movement, it is expensive, invasive and time consuming. 
It is also hard to reproduce the exact marker placement on different subjects and most methods assume that the markers are rigidly attached to the body, which is not true due to soft tissue motion.
MoSh \cite{Loper:SIGASIA:2014,mahmood2019amass} unifies mocap and statistical body models by fitting the parameters of the model to match the marker data.
This approach can even mitigate the issues of soft tissue motion.

Traditional mocap, however, typically prevents subjects from wearing normal clothing, complicating capture and limiting its applications.
Consequently, many research and commercial solutions for markerless motion capture exist \cite{uhlrich2022opencap, bittner2022camkinematics, peng2023skelpredictionnet}. 
For example, OpenCap \cite{uhlrich2022opencap} enables biomechanics from smartphone videos. They use OpenPose \cite{cao2017openpose} to detect the subject's 2D joint locations in several camera views and reconstruct their 3D locations. 
A biomechanical skeleton model is then fit to these 3D joints.
However, existing 2D joint detectors \cite{cao2017openpose, fang2022alphapose, mathis2018deeplabcut} have limited biomechanical accuracy, since they are typically trained using manually annotated 2D images.
The ``ground truth'' joint locations do not correspond to the %
actual functional 3D joint locations. 
When the 
joints predicted from images are compared to joint locations computed from motion capture systems ~\cite{needham2021jointsaccuracy}, the differences are as high as 30 to 50 mm for %
joints such as the knee.

{\bf Bones inside bodies.}
Our goal is to properly place the skeleton inside a parametric body model, providing the best of both worlds.
A common approach in previous work uses an anatomic skeleton model and deforms it to register it to a target body mesh \cite{Gilles2010, ali2013anatomy,saito2015computational, Zhu2015, Kadlecek2016}. 
This registration is challenging as these skeleton models do not, in contrast to SMPL, have a shape space of deformations. 
Thus the applied deformations may create non-plausible anatomies. 
In contrast, OSSO \cite{keller2022osso} learns to predict the geometry of the bones from a SMPL body mesh. 
They learn this 3D geometry from 2D medical images, where both the surface of the person and the skeleton can be observed. Although this approach gives a plausible skeleton shape that fits inside the subject, the resulting skeleton model can not be easily animated as it does not have a kinematic tree. 
For a lying pose, OSSO yields precise skeletal geometry that is close to the ground truth scans, but the reposing of the skeleton requires an optimization \changed{process} that can lead to biomechanically impossible poses. 
The recent BOSS model \cite{shetty2023boss} improves on OSSO by learning a skin-bone-organs model from segmented 3D medical data. While the skin and skeleton model share the same shape space, their kinematic trees used for rigging are different. This does not allow the synchronous posing of both skin and skeleton and an expensive optimization step is required. 

{\bf Bodies from bones.}
Going in the other direction, one can infer the body shape given a skeleton.
For example, BASH \cite{schleicher2021bash} uses the SCAPE body model \cite{anguelov2005scape} to envelop a biomechanical skeletal and muscle model \cite{nitschke2020efficient}.
However, the SCAPE model is only scaled to match the limb lengths of the skeleton.
Shape accuracy is not critical because their goal is to better visualize muscle activation by displaying it on the human surface.

In contrast to prior work, \skel provides a properly scaled skeleton inside any SMPL body model. 
Any optimization or regression method that estimates SMPL parameters can now be used to produce biomechanical skeletal parameters. 
\skel effectively connects parametric shape models with biomechanical skeletons for the first time to enable the integration of these technologies and fields.

\section{Method overview}

Our driving goal is to create \skel, a model that combines skin and skeleton meshes 
in which both are synchronously rigged with the same pose parameters $\qskel$,
and can be reshaped by inheriting the \smpl shape space.
To create this model, we must know the %
location of the anatomic joints and bone rotations inside the human body. 
There is no large-scale medical dataset of subjects in motion where one can extract both the body and skeleton meshes, and static medical scans do not fully constrain the skeleton in motion.
For this, we need bodies in motion and leverage the AMASS dataset \cite{mahmood2019amass} to address this challenge.
In \secref{sec:dataset} we first present our new custom \changed{Biomechanical Skeleton Model}, \osim, and describe how to align it inside AMASS sequences of \smpl bodies in motion to obtain the new \db dataset. 
Leveraging \db, \secref{sec:skel} shows how we learn the $\skel(\shape, \qskel)$ model, which inherits the shape space $\shape$ from \smpl and the pose vector $\qskel$ from the new \osim biomechanical model. It enables direct animation of the skin and the skeleton meshes using shape and pose parameters, $\shape$ and $\qskel$, respectively.
Creating \skel involves two important steps: \changed{learning the bone locations and orientations (\secref{sec:learning}) inside the body, and rigging the skin and bone motions to a common kinematic tree parameterized by $\qskel$ (\secref{sec:buildskel})}.

\section{The \db dataset}
\label{sec:dataset}

The goal of the \db dataset is to enable the learning of the location and orientation of the 3D bones inside a body surface in motion.
To create \db, we use the \smpl \cite{loper2015smpl} model for the body surface and a new biomechanical skeleton model, \osim, \changed{for the bones}. We first introduce \changed{these two models}, and then describe how we fit \osim to \smpl and create \db.

\subsection{The SMPL body surface model} 
We model the 3D body surface using the \smpl function, which takes as input shape parameters $\shape$ and pose parameters $\pose \in \mathbf{R}^{\smplNbdof}$, and outputs a 3D mesh with vertices $\vect{v} \in \mathbf{R}^{6890 \times 3}$.
The \smpl model includes a joint regressor defined in Eq.~10 of \cite{loper2015smpl}. %
It computes the 3D joint locations of the kinematic tree for the shape parameters $\shape$.
Each joint is parameterized by three degrees of freedom in an axis-angle representation.
The \smpl kinematic tree is artist-defined and only approximately corresponds to the human anatomy.
The \smpl equation, summarized in Eq.~5 and 6 of \cite{loper2015smpl},
first deforms a template mesh $\smpltemplate$ using learned deformations driven by the shape and pose parameters. Then linear blend skinning (LBS) is used to pose the vertices and produce the body mesh vertices.

\subsection{The \osim skeletal model} 
\label{sec:osim}

To model the human skeleton, we create \osim, a custom skeleton model using the OpenSim framework \cite{delp2007opensim}; \changed{\osim is described by a file in~``.osim'' format}.
\osim consists of $\osimNbparts$ rigid groups of bones with joints defined between them as well as a mesh representing the geometry of each bone group.
On top of each bone, a set of virtual markers is defined; these markers are used to fit \osim to motion capture sequences.

The $\osim$ is represented by three functions that take scaling and pose parameters as input.
\changed{Using forward kinematics}, these functions
output
the skeleton joint locations,  $\osim^J(\scales, \qosim)$, 
the bone meshes vertices, $\osim^v(\scales, \qosim)$, and the posed marker locations,  $\osim^m(\scales, \qosim, \mosimtemplate)$. 
The scale parameter $\scales \in \mathbb{R}^{\osimNbparts \times 3}$ scales \changed{each of the \osimNbparts unposed} bones along the axis (x,y,z), \changed{while} 
the pose parameters $\qosim \in \mathbb{R}^{\osimNbdof}$ represent the \osimNbdof degrees of freedom of the articulated model.
The model markers are defined by designating their 3D coordinates \changed{$\mosimtemplate \in \mathbb{R}^{N_m\times3}$} in the corresponding bone reference frame. Each marker is rigidly attached to one bone and, when the bone is scaled with $\scales$, the marker location is scaled accordingly. In contrast to \smpl, \osim has a more realistic kinematic tree but lacks a shape space. 

\moved{[Moved in from Sup.~Mat.]}
\changed{Body models like SMPL typically treat every joint as a ball joint with three angular degrees of freedom.
In reality, the joints of the body differ significantly from this assumption. 
Consequently, for \osim we use more realistic models of the spine, shoulder, and forearm.
}

{\bf Lower body.} \changed{For \osim's lower body model, we use the model from Rajagopal et al.~\shortcite{rajagopal2016full}, which implements the knee flexion model from Walker et al.~\shortcite{walker1988knee}.}

{\bf Spine.} We extend the original OpenSim framework with a new custom joint that we call ``constant curvature'', to model the spine bending with a constant length. Our \osim model's spine is made of 3 such joints, enabling lumbar, thoracic, and cervical bending, as illustrated in \figref{fig:spine}.
Given the parent joint location $J_{i-1}$ and a spine curve of length $l$, the child joint $J_i$ will move on a curve of constant arc length and curvature, parameterized by one termination angle $\qosim_i = [q_{x}, q_{z}, q_{y}] \in \mathbb{R}^3$, represented as Euler-angles in XZY.
The child joint location is $J_{i} = R(\qosim_i) \cdot (J_{i-1} - J_{i}) + \vect{t}^{\text{spine}}(\qosim_i)$, where 
\begin{equation}
\begin{aligned}
&\vect{t}^\text{spine}(\alpha) = \left\{\begin{matrix}
x =& (r - \cos(\alpha)) * \frac{-sin(q_{z})}{sin(\alpha)}\\ 
y =& r \cdot \sin(\alpha) \\
z =& (r - \cos(\alpha)) * \frac{cos(q_{z}) * sin(q_{x})}{sin(\alpha)}
\end{matrix}\right. \\
&\text{with~} \alpha = \arcsin{\sqrt{sin(q_{z})^2 + (cos(q_{z}) * sin(q_{x}))^2}}, \text{~and~} r = \frac{l}{\alpha}. 
\end{aligned}
\label{eq:spine}
\end{equation}

{\bf Shoulder blades.} In \osim we follow the \citet{seth2016scapulothoracic} model and parameterize the shoulder blade joint such that it slides along an ellipsoid defined around the thorax, making the scapula slide along the ribs. The three degrees of freedom are linked to scapula abduction, elevation, and upward rotation as illustrated in \figref{fig:scapula}.

\textbf{Forearm.} The forearm pronation and supination are modeled by a single degree of freedom; which is distinct from the elbow flexion, wrist flexion, and wrist deviation \cite{rajagopal2016full}.
The forearm is made of two bones: the radius and the ulna. The ulna is linked to the humerus through a hinge joint, enabling the elbow flexion. During the forearm pronation and supination, the hand rotates while the ulna stays fixed. We model this by rotating the radius along the axis defined by the ulna's parent joint location and the radius extremity as illustrated in \figref{fig:forearm}. \moved{[End moved material.]}

\subsection{Fitting \osim to \smpl}
\label{sec:bsm_fit}
\begin{figure}[t]
    \centering
    \begin{subfigure}[b]{0.3\columnwidth}
        \centering
        \includegraphics[width=\textwidth]{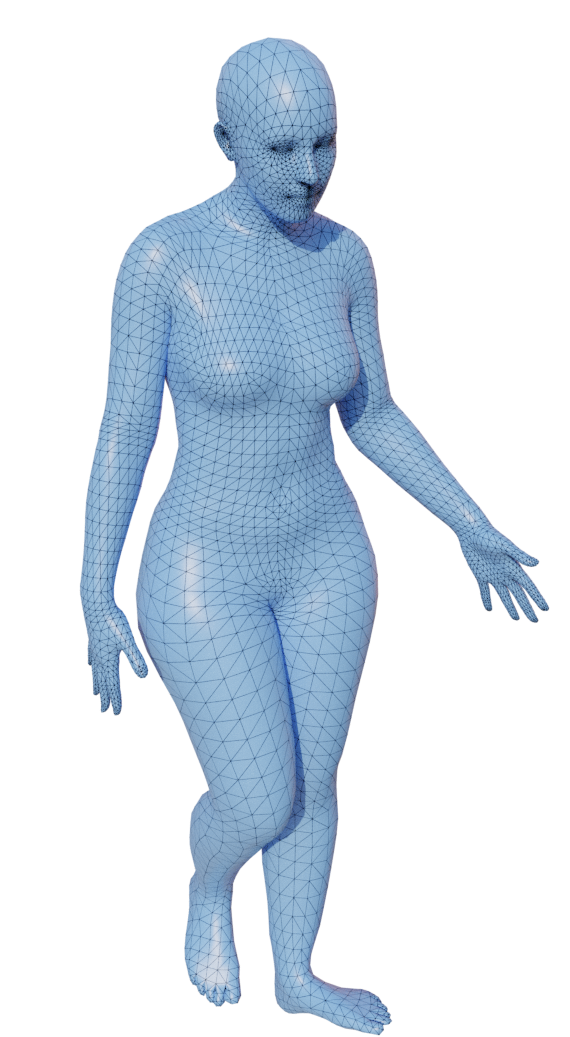}
        \caption{\smpl}
        \label{fig:amasssmpl}
    \end{subfigure}
    \begin{subfigure}[b]{0.3\columnwidth}
        \centering
        \includegraphics[width=\textwidth]{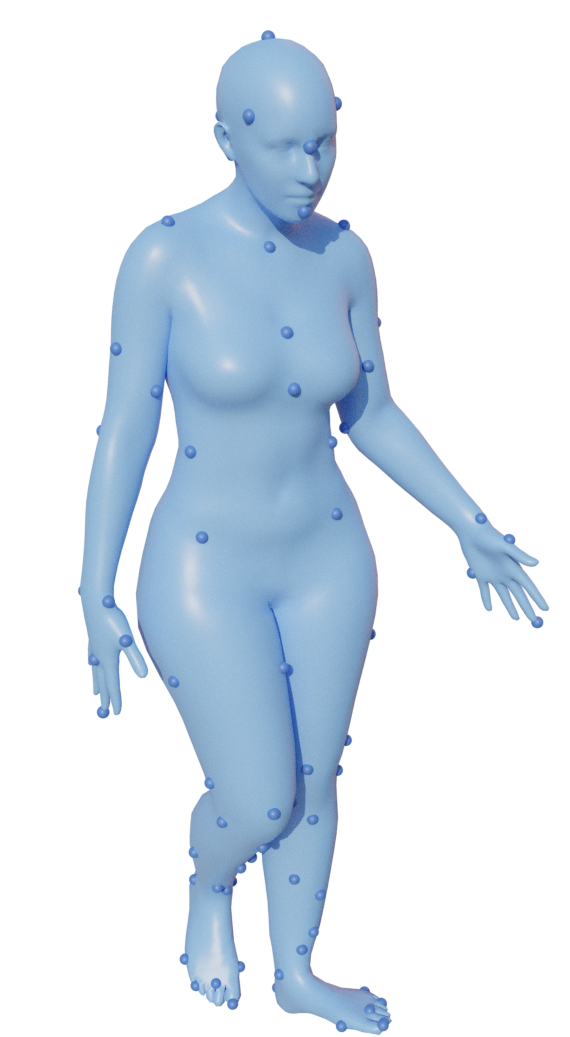}
        \caption{Synthetic markers}
        \label{fig:amasssynthmarkers}
    \end{subfigure}
    \begin{subfigure}[b]{0.3\columnwidth}
        \centering
        \includegraphics[width=\textwidth]{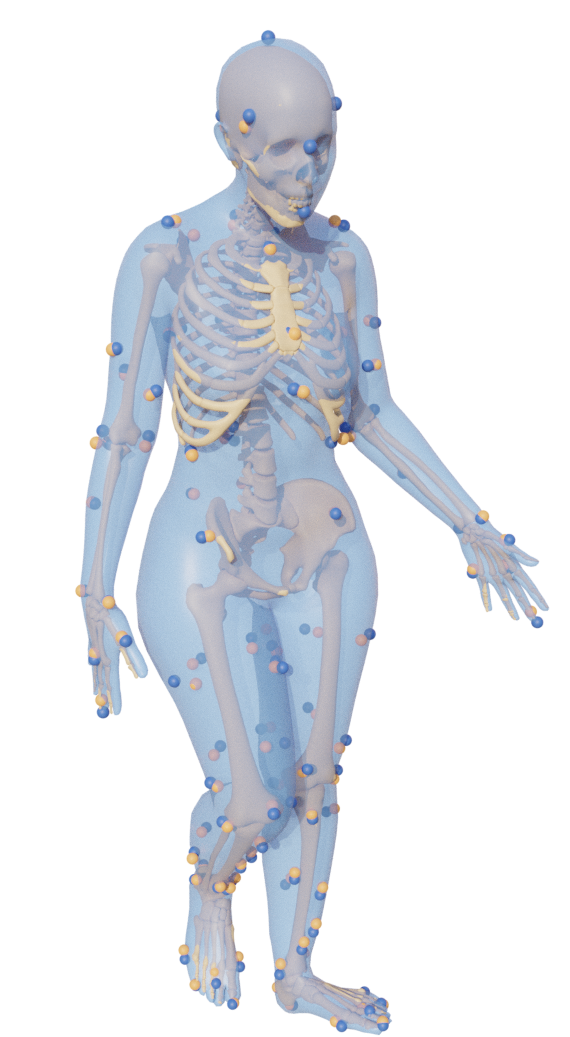}
        \caption{\osim alignment}
        \label{fig:amassosim}
    \end{subfigure}
       \caption{Creation of the paired skeleton and body dataset. Given a \smpl motion sequence (a), we generate synthetic markers (b), and \changed{fit} a biomechanical model to \changed{the makers} using \ab \cite{werling2022addbio} (c).}
   \label{fig:creation}
\end{figure}

To leverage the \smpl body meshes in AMASS, we define a mocap marker set on the \smpl mesh and obtain synthetic sequences of markers.
We use these as input to fit our \osim skeleton using \ab \cite{werling2022addbio}, a \changed{recent} biomechanical optimization framework.
\figref{fig:creation} illustrates this pipeline.

\subsubsection{Establishing marker correspondences with \osim}
\label{sec:osso_placement}
\changed{To fit \osim to \smpl, we define the same markers on both models. Theoretically, we could define each skin vertex of SMPL to be a marker attached to \osim. But OpenSim rigidly attaches markers to the bones, hence we define a set of markers that are mostly influenced by one bone and not subject to significant soft tissue deformation}. 

\begin{figure}[t]
    \centering
    \includegraphics[width=\columnwidth]{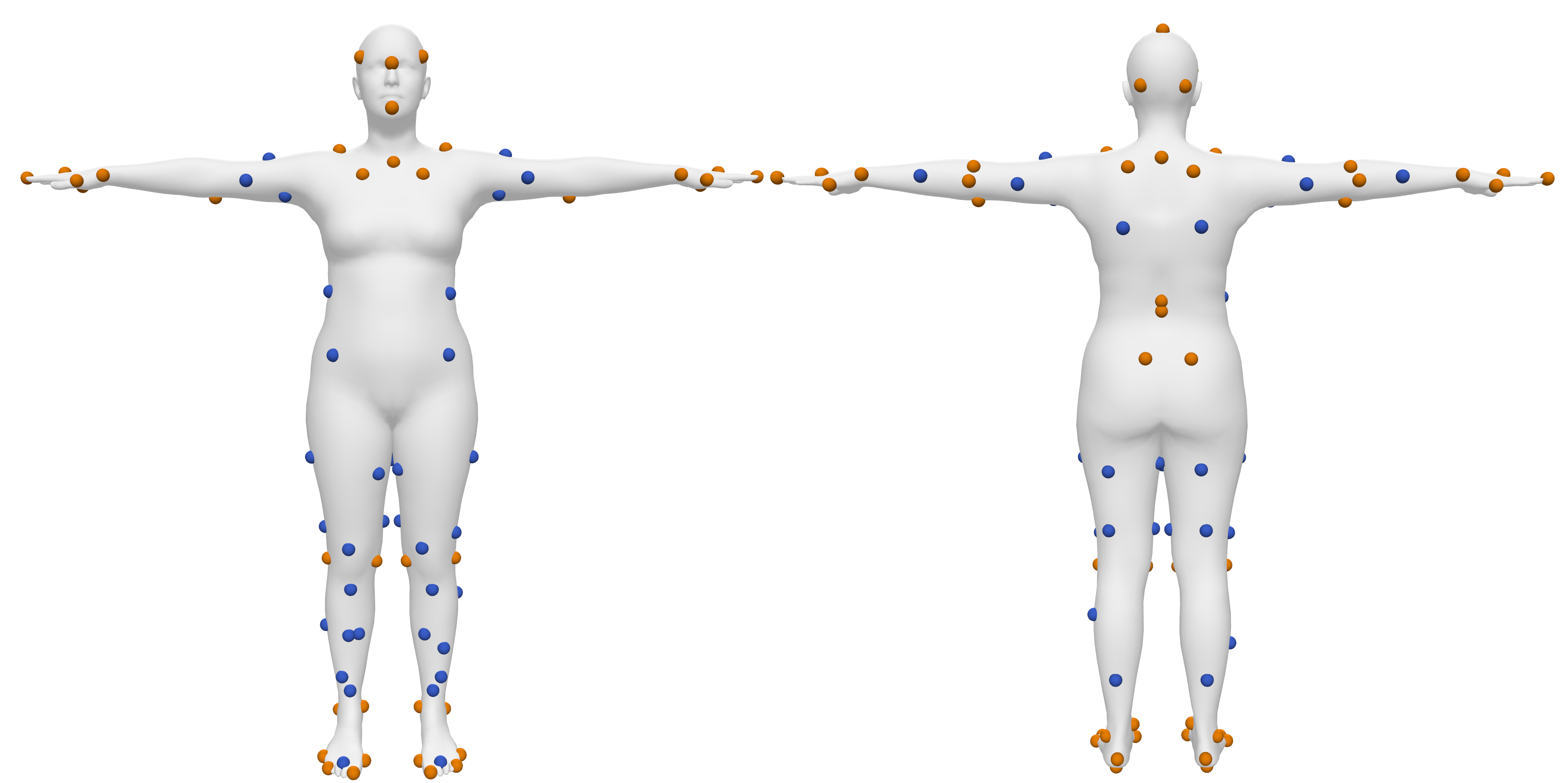} 
    \caption{\changed{The markers defined on SMPL: bony in orange, soft in blue.}}
    \label{fig:all_markers}
\end{figure}

Specifically, we define $\NbBonyLdm$ \textit{bony markers} that are close to the bones, \changed{as typically done in motion capture}.
\changed{Each marker is defined on \osim and \smpl by examining tight \smpl fits to 3D scans and identifying specific SMPL vertices. Figure \ref{fig:all_markers} shows all the \textit{bony markers} on \smpl in orange.}

\changed{Although this marker set follows the rigidity assumption, is too sparse in some areas to properly constrain the location of the bones. So we introduce an additional $\NbSoftLdm$ \textit{soft markers}, located on soft body parts (blue in \figref{fig:all_markers})}.
\changed{To define a new \osim marker, it needs to be positioned on the \osim skeleton template. While this can be achieved quite precisely for \textit{bony markers}, it is harder to estimate at what distance to the bones \textit{soft markers} should lie. Moreover, this distance varies significantly for different body shapes (e.g.~due to adipose tissue). Initializing markers close to the bones for subjects with more adipose tissue can lead to \ab over-stretching the bones to fit the SMPL markers.}

\changed{To address this marker offset issue, we propose a method to automatically define markers on \osim with personalized offsets depending on the body shape.}
We leverage the \osso model~\cite{keller2022osso}, which predicts the location and shape of the skeleton inside \smpl. \changed{In contrast to \osim, OSSO models the geometry of the skeleton with respect to the body shape and, as it was trained on medical scans, it learned the offset between the bones and the skin. We can thus use it to compute where skin markers should be located with respect to the bone surface, given a body shape.}
We first compute the relationship between the \osso and \osim bones. Precisely, we register each \osso bone mesh to the corresponding \osim bone mesh and effectively obtain all \osso bones in the reference frames of the \osim bones. This relationship only needs to be computed once.
Then, for each AMASS subject, we use \osso to obtain their skeleton mesh. We use the lying down pose in which \osso is trained to obtain the best possible \osso prediction.
Now, given a marker location on the \smpl mesh and the computed \osso bone mesh inside the body, we parameterize the marker location using the closest triangle on the \osso bone mesh (\figref{fig:osso_markers}). %
This allows us to transfer the marker location onto the \osso bone mesh and, consequently to the corresponding template \osim bone (\figref{fig:osim_markers}).

\changed{We use this method to generate a \osim model for each subject, with personalized markers $\mosimtemplate(\shape)$, thus avoiding over-stretching the bones during the \ab optimization, as shown \figref{fig:osso_osim_markers_comparison}.}
\changed{We experimented with different marker sets, adjusting their number and placement, to obtain the best possible fits from \ab; i.e.~minimizing the marker errors and yielding a satisfactory fit visually.}

\subsubsection{Fitting \osim to motion data}
With corresponding markers defined on both \smpl and \osim, we can fit the \osim skeleton to any \smpl mesh. 
Given a sequence of $N_f$ %
frames and $N_m$ %
target 3D marker locations per frame, ${\targetmarkers_k}$ $(k \in \{1,\dots, N_m\})$, extracted from the sequence of \smpl meshes, we use \ab \cite{werling2022addbio} to obtain the \osim scale parameters $\scales$ and the $N_f$ poses $\{\qosim_f\}$. We optimize a bi-level objective, to find the best $\scales$ such that inverse-kinematics with these scales yields poses $\{\qosim_f\}$ with minimal distance to the $N_m$ target markers:
{\small
\begin{equation}
    \argmin{\scales, \markeroffset} \bigg( \big( \sum_{f=1}^{N_f} \argmin{\qosim_f}  \sum_{k=1}^{N_m} \lambda_k (\mathrm{\osim}^m(\scales, \qosim_f, \mosimtemplate(\shape)+\markeroffset)_k - \targetmarkers_k) \big) + \lambda_{p} P(\scales, \shape) \bigg),
\end{equation}
}
where $\markeroffset \in \mathbb{R}^{N_m\times 3}$ is a 3D per marker offset. 
The weighting factor $\lambda_k \in \mathbb{R}$ is set to a low value for {\it soft markers} and a high value for {\it bony markers} to allow larger fitting errors due to secondary soft tissue motions.

\changed{The prior $P$ regularizes the scale of the bones, given the subject's height, weight, and biological sex as in \cite{werling2022addbio}. We automatically estimate the height and weight of each subject from their \smpl shape parameters $\shape$, by assuming that the body has a uniform density \cite{pujades2019virtual, muller2022shapy} and thus re-parameterize this prior term as $P(\scales, \shape)$.}

\begin{figure}[t]
    \centering
        \begin{subfigure}[]{0.3\columnwidth}
        \centering
        \includegraphics[height=3.5cm]{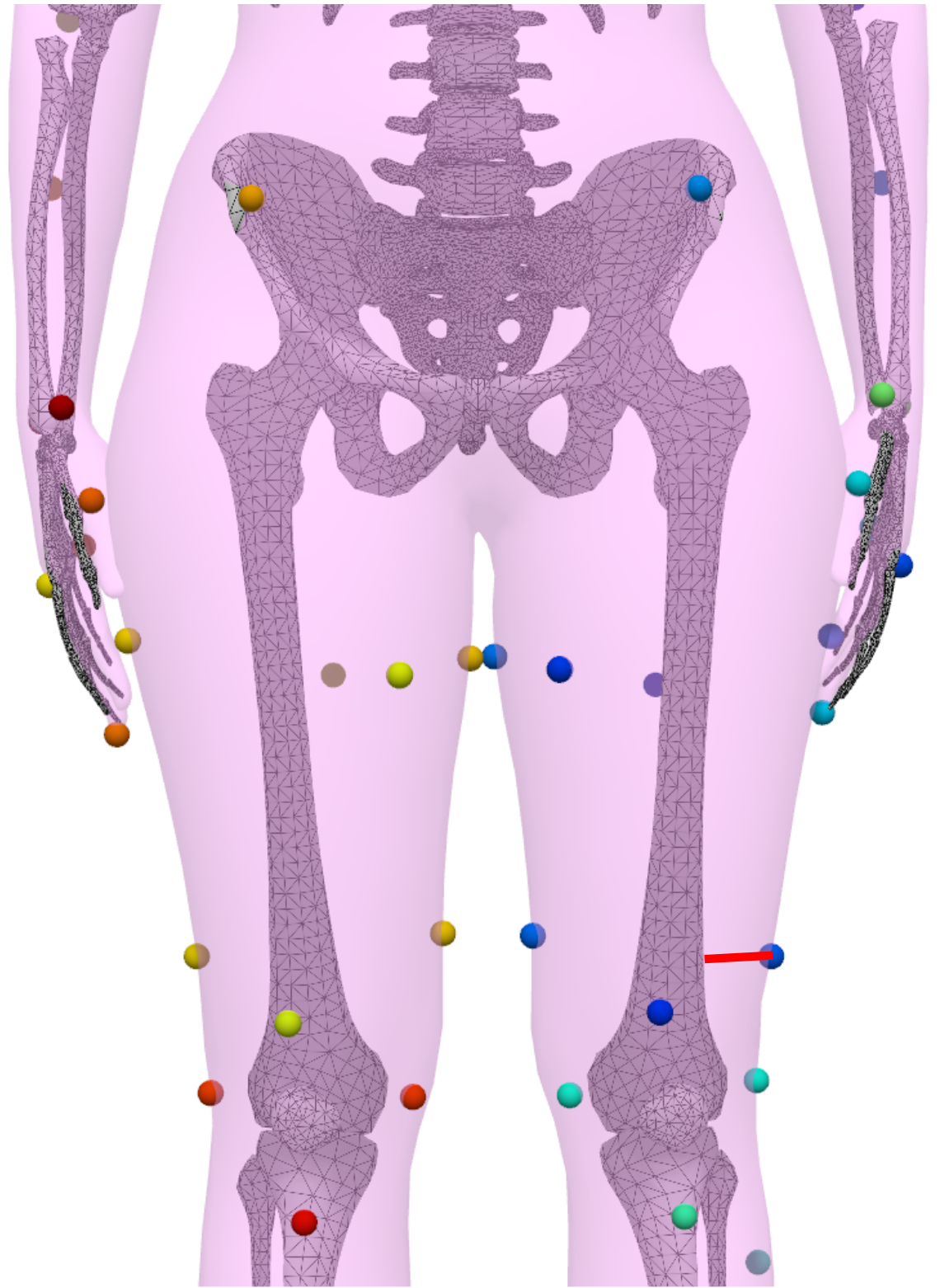}
        \caption{Markers wrt OSSO}
        \label{fig:osso_markers}
    \end{subfigure}
    \hspace{0.5cm}
    \begin{subfigure}[]{0.2\columnwidth}
        \centering
        \includegraphics[height=4cm]{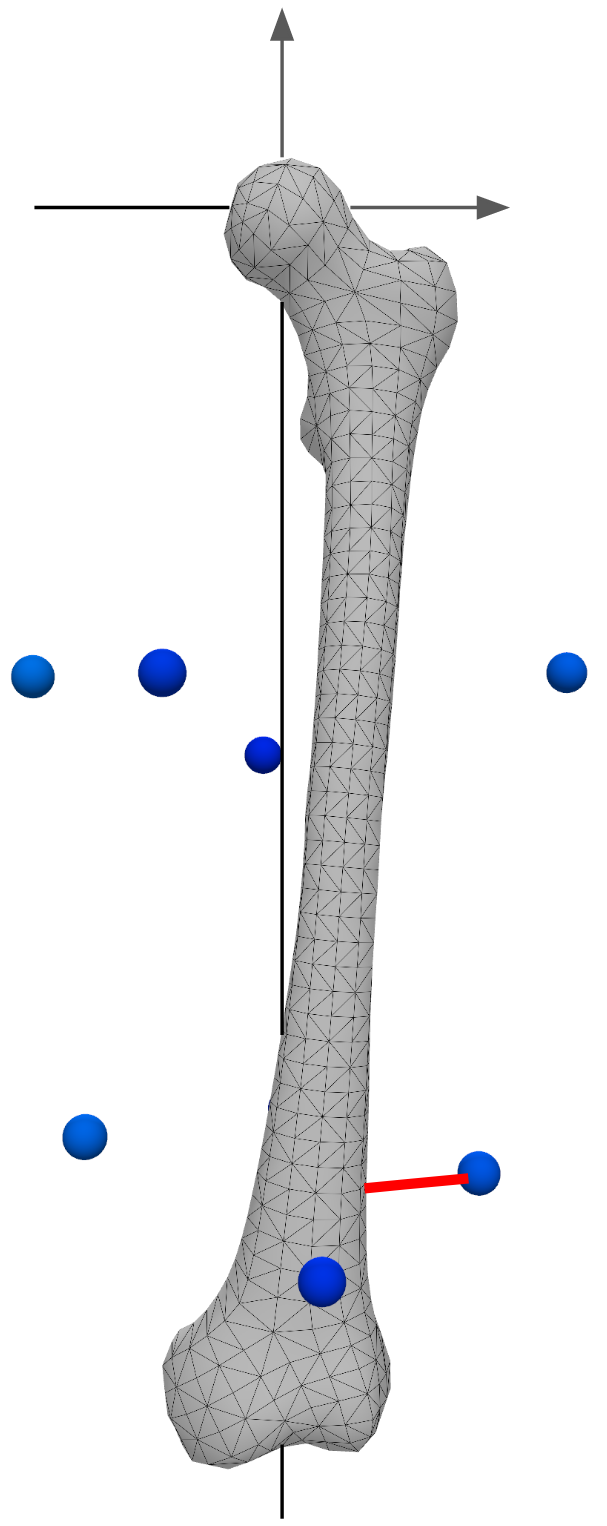}
        \caption{Markers in the \osim bone frame}
        \label{fig:osim_markers}
    \end{subfigure}
    \hspace{0.5cm}
    \begin{subfigure}[]{0.3\columnwidth}
        \centering
        \includegraphics[height=4cm]{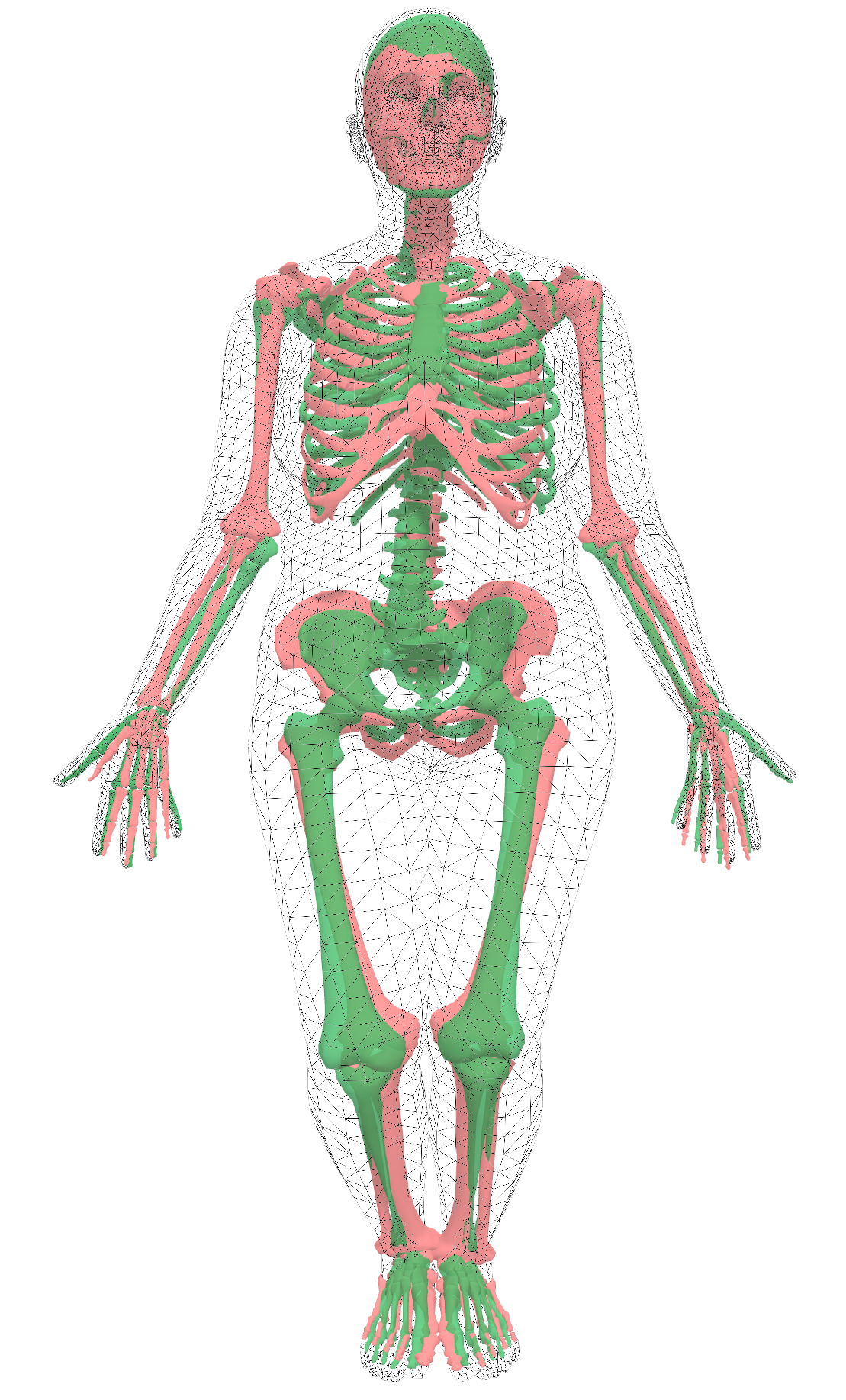}
        \caption{\ab fit result}
        \label{fig:osso_osim_markers_comparison}
    \end{subfigure}
    \caption{\changed{\textbf{(a)} The OSSO skeleton is aligned to the subject's SMPL mesh. \textbf{(b)} We deduce the personalized markers location $\mosimtemplate(\shape)$ on the \osim bone template.
    \textbf{(c)} On high BMI subjects, a shape-agnostic marker definition for all subjects yields over-stretched bones (\textbf{red}). Using personalized marker locations $\mosimtemplate(\shape)$ defined using OSSO prevents this over-stretching (\textbf{green}).}}
    \label{fig:marker_transfer}
\end{figure}

Despite the scale prior, using a generic marker set can lead to \ab over-stretching the bones for heavy subjects.
Defining personalized marker locations $\mosimtemplate(\shape)$ on the skeleton template as described in the previous section helps further regularize the bone scales (\figref{fig:osso_osim_markers_comparison}).

We apply this \changed{optimization process} to a subset of AMASS consisting of 113 subjects and 2198 motion sequences, amounting to over 9 hours of motion data.
The paired \smpl meshes and \osim skeletons form the \db dataset.
For each subject $p$ there is a \smpl body shape $\shape_p$
and the scaled personalized \osim model $\scales_p$.
Further, for each motion frame $f$ it includes the bone angles $\qosim_f$ as well as the bone joint locations $\josim_f$.
Figure \ref{fig:osim_align} shows examples of the \db dataset.

\def\w{0.30\columnwidth} 
\newcommand{\osimrender}[2]{
  \includegraphics[width=\w,trim={0cm #1 0cm 0cm},clip]{figures/scripted/frame_rendering/#2_smpl_osim.png}
}
\begin{figure}[b]
    \centering
    \osimrender{200}{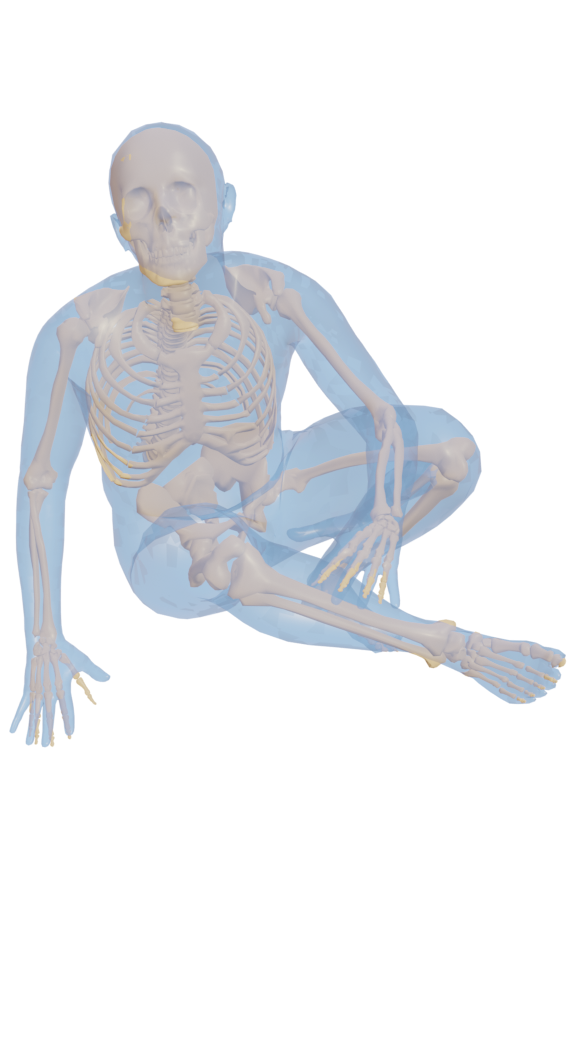} %
    \osimrender{0}{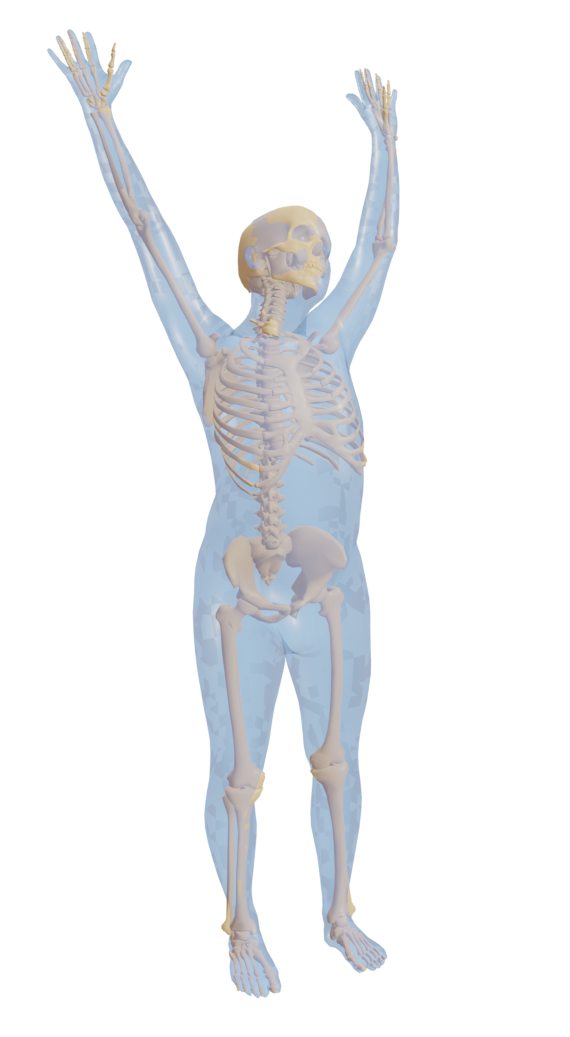}
    \osimrender{0}{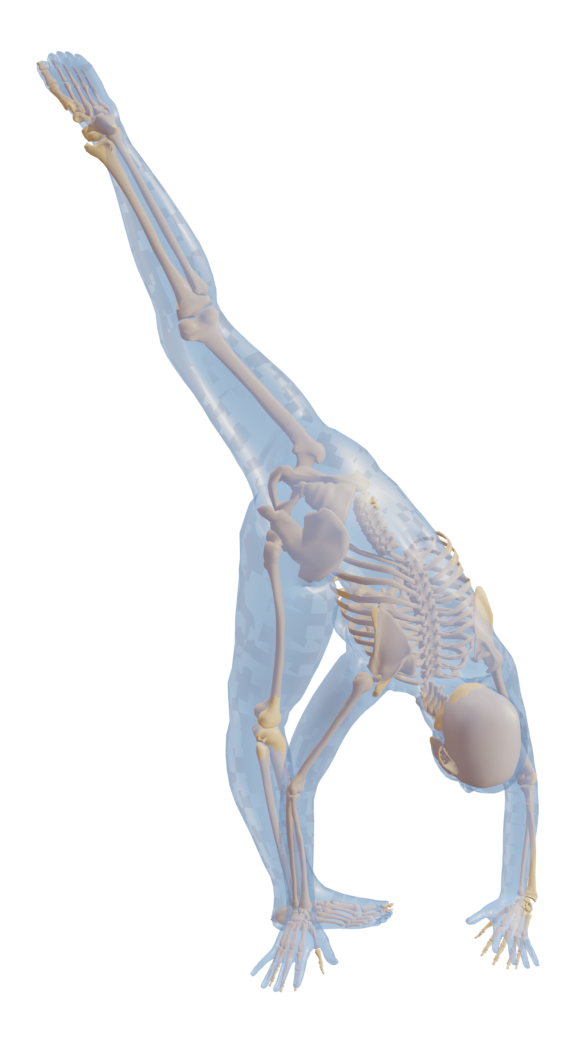}
    \caption{\db: examples of \osim fits to AMASS poses.}
    \label{fig:osim_align}
\end{figure}

\section{The SKEL model}
\label{sec:skel}

\changed{Now we have \osim skeletons inside \smpl but we want to go further and parameterize the 3D body model with the biomechanical skeleton.}
\changed{To that end, we develop} \skel, which is designed to be compatible with \smpl and \changed{posed like} \osim. 
This allows us to leverage \smpl's learned shape space as well as all the existing datasets where \smpl bodies are estimated from different modalities.
To create \skel, we must put \smpl vertices and the \osim skeleton together in the same reference frame.
We pose the \osim skeleton mesh inside a \smpl body in T-pose, the zero-pose of the \smpl body model.
To that end, in \secref{sec:learning}, we learn to regress the anatomical joint locations from \smpl using the \db dataset.
Then, in \secref{sec:buildskel} we describe how we rig the \smpl model using the \osim joint rotations.

\begin{figure}[t]
\centerline{
    \includegraphics[width=\columnwidth]{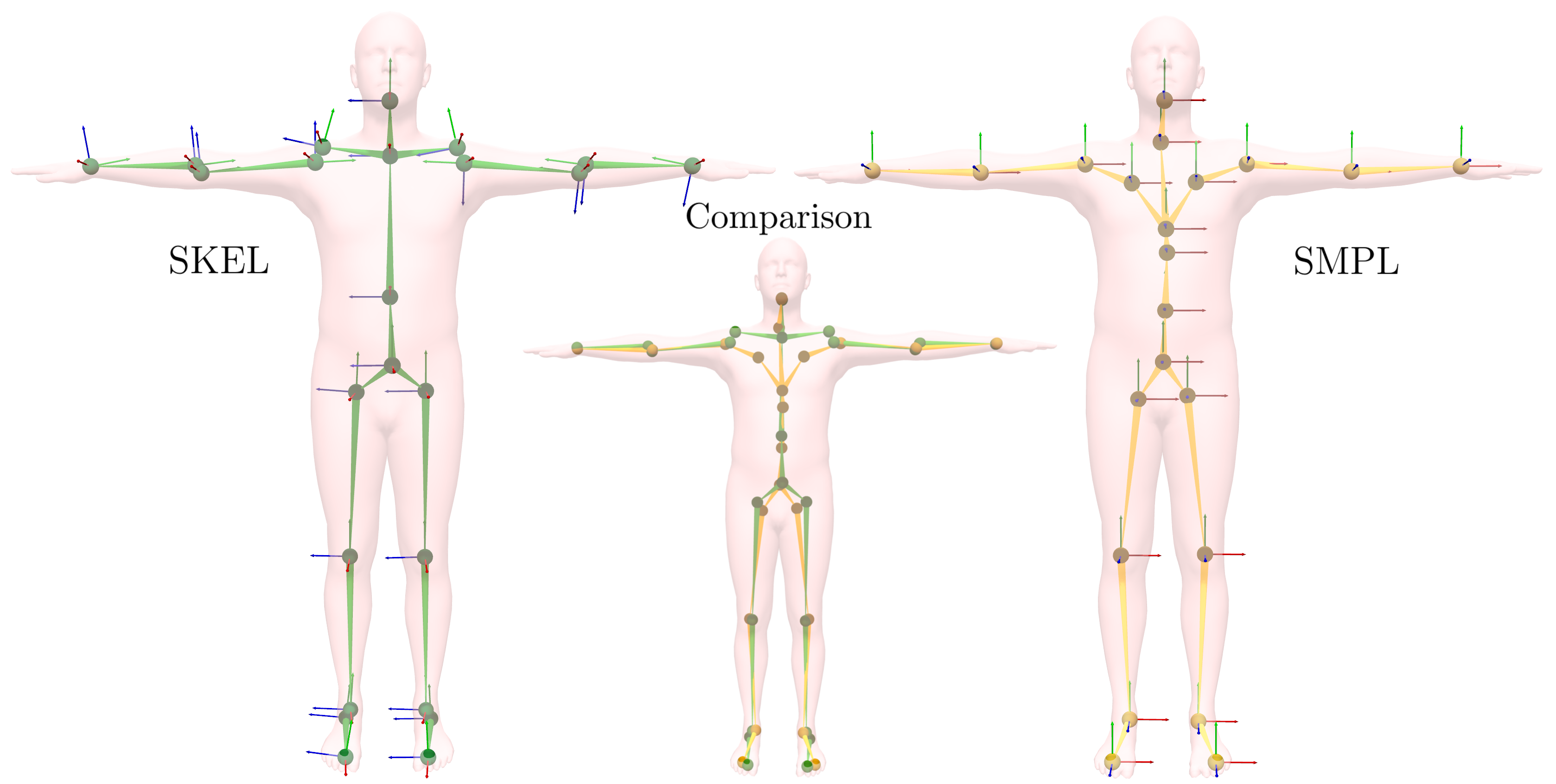}
}
   \caption{\textbf{Left:} \skel kinematic tree with learned anatomical joint locations. \textbf{Right:} \smpl's kinematic tree. \textbf{Middle:} the superposition of both. \changed{In contrast} to \smpl, which has axis-aligned rotation axes, \skel's rotation axes are bone-aligned. 
   }
    \label{fig:kintree}
\end{figure}

Note that, in \smpl, all joint orientations are defined in a global T-pose space with an axis-aligned frame of reference for each joint as illustrated in \figref{fig:kintree} right. This means that \smpl assumes, for example, that the elbow rotation axis is aligned with the world y-axis, independent of the orientation of the humerus. The overparametrized nature of SMPL allows plausible arm articulation by combining several axis rotations. 
But \osim, with its reduced degrees of freedom for the rotations, requires the local frame on which the rotation is applied to be precisely aligned with the anatomy
in order to obtain a proper anatomic rigging.
In addition, the location and orientation of the humerus and ulna bones have to be coherent with the rotation axis.
In \smpl this coherence does not exist: the joint reference frames are not aligned with the articulation axis. As shown in \figref{fig:kintree}, the elbow frame is not aligned with the segment defining the humerus position.
Hence, we first learn to predict the location of the joints inside \smpl and, with these, we learn to properly orient the bones inside a \smpl body mesh.

\subsection{Establishing the bone locations and orientations}
\label{sec:learning}

\paragraph{Anatomical joint locations}
Given paired \smpl body meshes and their corresponding \osim anatomic joint locations, we learn a function that predicts the joints from the body surface.
We proceed similarly to \citet{loper2015smpl} by learning a joint regressor $\jointreg$ that takes as input the \smpl mesh vertices $\smplverts \in \mathbb{R}^{6890 \times 3}$ and predicts the new anatomic joints $\osimjoints \in \mathbb{R}^{\osimNbparts \times 3}$.
We follow the \citet{keller2022osso} \osso methodology, 
by formulating a non-negative least squares problem for each joint $i$, and solving it with an active set method \cite{lawson1995solving}.
\changed{We train these regressors from the posed vertices and joints of the \db dataset.}

Figure \ref{fig:kintree} shows the new regressed kinematic tree in green.
Notice that the hip joint locations, corresponding to the femur heads, are more anatomically correct than the ones in \smpl.
The comparison also shows significant differences at the shoulders,
as well as more subtle, but important, differences for the other joints.

\paragraph{Bone orientations}
\changed{We aim to find the orientation of the bones inside the \smpl T-pose mesh, i.e. find the rotation $R_i$ to apply to the i-th \osim bone template mesh, to position it inside the \smpl T-pose mesh.}
In \osim, the rest position of each individual bone template is centered at the origin and oriented along the canonical axis x, y, z. In the following, we refer to the ``bone axis'' as the axis passing through the bone's proximal and distal ends.

\changed{In contrast to \osim, in \smpl T-pose, the bones should be positioned and oriented between pairs of regressed anatomical joints. This brings two challenges: (i) the rotation of the bone around its bone axis is not known, and (ii) as the regressed joint location depends on $\shape$, the orientation of the bones also varies with $\shape$.}

\changed{To solve those two issues, we split the bone rotation $R_i(\shape)$ into a learned base rotation $R_i^{base}$ and a shape-dependant rotation $R_i^{\shape}(\shape)$:}
\begin{equation}
    R_i(\shape) = R_i^{\shape}(\shape) \cdot R_i^{base},
    \label{eq:bone_tpose_rot}
\end{equation}
\changed{where $R_i^{base}$ is learned to define the bone's orientation around its bone axis, ensuring that bones are properly orientated wrt their parent bone. $R_i^{\shape}(\shape)$ is computed dynamically to align the bone to the segment defined by its parent and child joints, so that the bone stays in its socket regardless of the shape of the subject}.

First, we learn $R_i^{base}$ from \db.
For each bone $i$, we can define a corresponding \smpl joint and limb. 
For example, the right humerus bone corresponds to the 17th joint and the right upper arm of \smpl. 
Thus, for each frame $f$ of our dataset, we obtain the bone \osim rotation $R_{i,f}^\osimexponent$ and its \smpl rotation $R_{i,f}^S$.

$R_i^{base}$ is the rotation that the bone needs to undergo so that when chained to the \smpl rotation, the corresponding \osim rotation is obtained. 
For each bone $i$ we learn its base rotation $R_i^{base}$ by minimizing
\begin{equation}
\sum_{f=1}^{N_F} (R_{i,f}^\osimexponent - R_{i,f}^S R_i^{base})^2
\end{equation}
over the $N_F$ frames of the dataset.

\begin{figure}[t]
\centering
    \includegraphics[width=0.8\columnwidth]{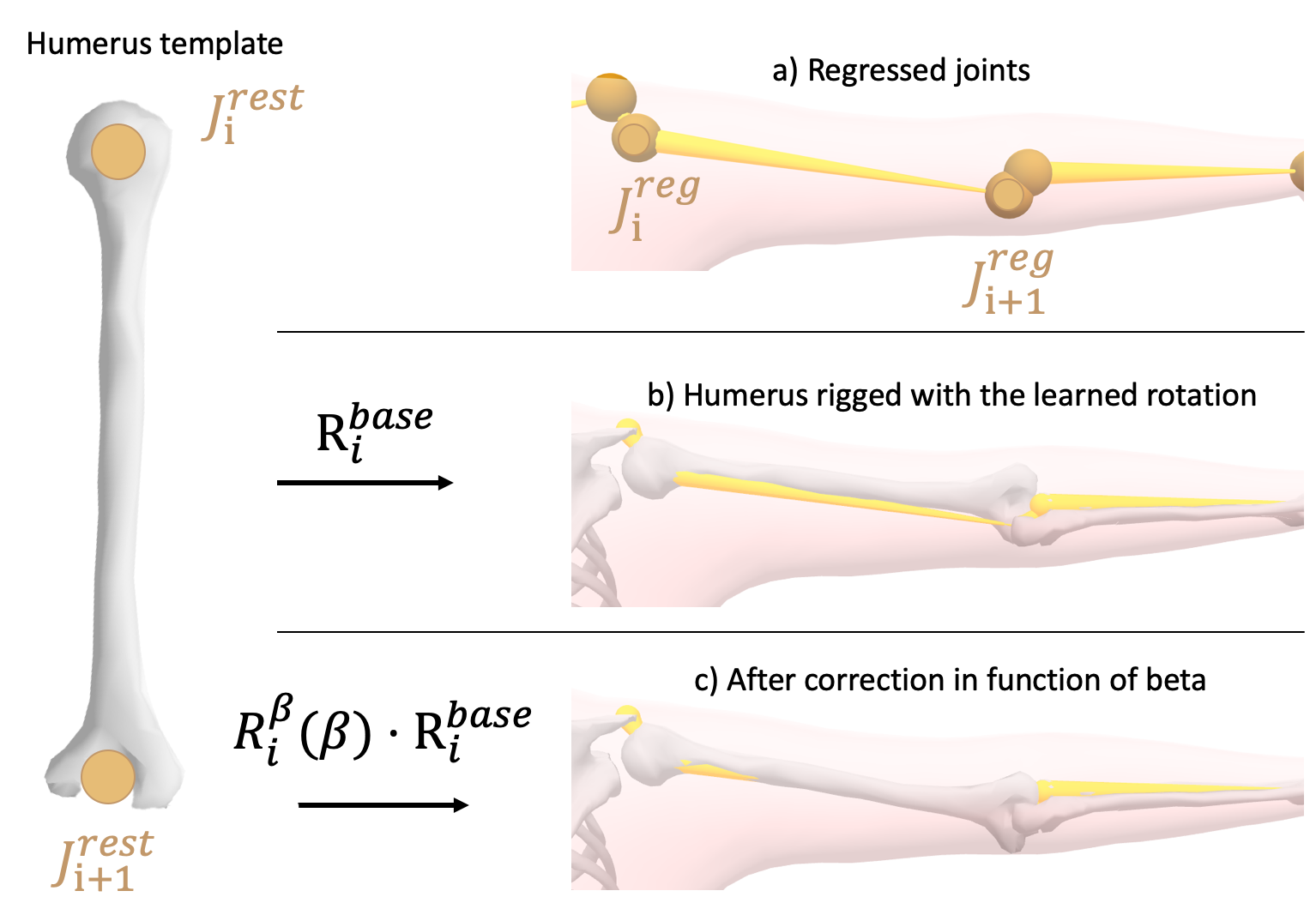}
   \caption{
   Left: Humerus template in the rest pose. We want to find its transformation to position it inside \smpl's arm.
    On the right we show, in order: a) the anatomical bone joints $J^{reg}_i$ regressed from \smpl skin vertices (pink). b) We center the bone on $J^{reg}_i$ and orient it with $R_i^{base}$. \changed{This provides a rough alignment, rotating the bone properly around its bone axis.}
    c) We then compute and apply the personalized rotation $R_i^{\shape}(\shape)$ to perfectly align the bone with the limb segment. Notice how the ulnar head now properly fits in the humerus distal end.}
   \label{fig:bonealign}
\end{figure}

\changed{This rotation properly orients the bone around its bone axis. But, as shown in \figref{fig:bonealign}, this rotation alone does not guarantee that the bones are aligned between their T-pose parent and child joints.}

Thus we explicitly compute $R_i^{\shape}(\shape)$, a shape-dependent corrective rotation 
that aligns the bone segment $(R_i^{base}~(J^{rest}_{i+1} - J^{rest}_i))$
with $(J^{reg}_{i+1}(\shape) - J^{reg}_i(\shape))$,
 where $J^{rest}_i$ is the location of joint $i$ in the bone rest pose 
 and $J^{reg}_i(\shape)$ is the shape-dependent regressed joint.
 The rotation axis of $R_i^{\shape}(\shape)$ is computed \changed{from} the cross-product of the segments.
As shown in \figref{fig:bonealign}, this effectively ensures a proper fit of the bone geometry into the regressed joint location.

It is worth noting that computing a direct rotation between the rest bone and the regressed segment 
$(J^{reg}_{i+1}(\shape) - J^{reg}_i(\shape))$ leaves a degree of rotation open: the rotation around the bone axis.
With the proposed approach, we obtain an anatomically coherent placement of the skeleton. Thanks to \db, a consensus orientation $R_i^{base}$ is found, which is then specialized per subject with $R_i^{\shape}(\shape)$. 
Effectively, the per-joint $R_i^{base}$ is learned from the dataset once and kept fixed,
and each per-joint $R_i^{\shape}(\shape)$ is recomputed when the shape parameters change.

\subsection{Building \skel: A single rig for skin and bones}
\label{sec:buildskel}

\begin{figure}[t]
    \centering
    \includegraphics[width=0.8\columnwidth]{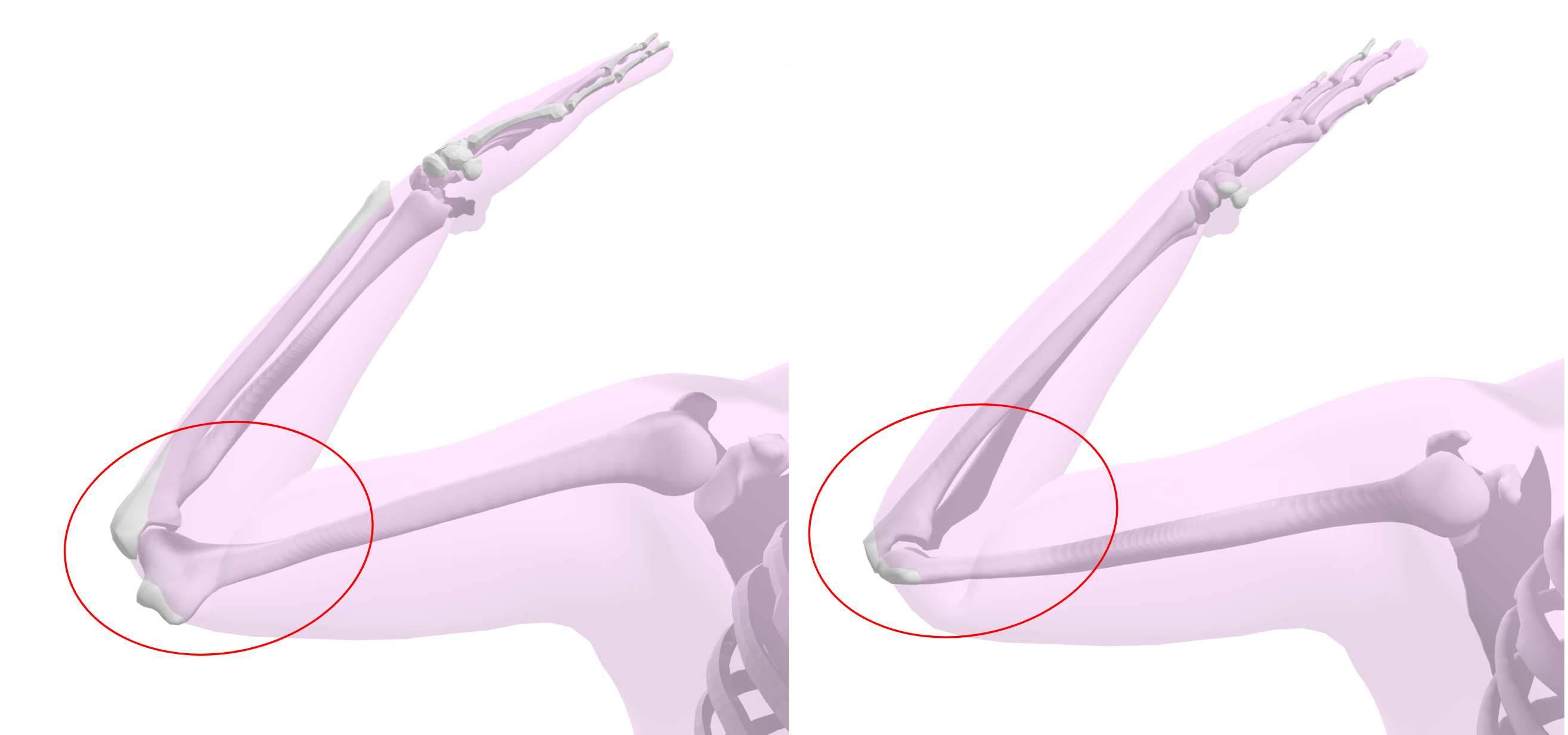}
    \caption{Left: Rigging the skeleton to the regressed joints and posing them using SMPL parameters $\pose$ can yield unrealistic articulations. We see that the humerus posed with the SMPL upper arm transformation does not yield the correct humerus orientation. Right: \osim fit for the same frame.}
    \label{fig:bad_smpl_rigging}
\end{figure}

As we saw in \figref{fig:kintree}, the \smpl kinematic tree is not suited to rig the skeletal structure, as its joints do not match the anatomic ones. \changed{Moreover, because of its over-parameterization, applying SMPL's transformation to the bones can yield unrealistic bone orientations, as shown in \figref{fig:bad_smpl_rigging}}.

Consequently, we re-rig \smpl with new anatomic degrees of freedom using the learned bone locations and orientations (\secref{sec:learning}).

\paragraph{The \skel function}
The \skel function takes as input a vector of \smpl shape parameters, $\shape$,
and the $\qosim \in \mathbb{R}^{\osimNbdof}$ pose parameters of \osim.
\skel outputs $(\skelskinverts, \skelskelverts, \jskel)$ where $\skelskinverts$ are the body surface vertices, $\skelskelverts$ the skeleton mesh vertices, and $\jskel$ the learned anatomic joint locations.

{\bf Skin.} \skel builds on the additive approach of \smpl, starting with a mean template mesh $\smpltemplate \in \mathbb{R}^{6890 \times 3}$ and adding the learned displacements $\shape \cdot \shapespace + \posedirs(\qskel)$, where $\shapespace$ is the PCA shape basis learned in \citet{loper2015smpl}
and $\posedirs(\qskel)$ are pose dependent displacements.
The posed \skel body vertices $\skelskinverts$ are then computed \changed{with the following linear blend skinning equation:} 
\begin{equation}
\skelskinverts(\shape, \qskel)
=
\left[
\changed{\sum_{i=1}^{\osimNj}}W_{i}^{\text{skin}} G_i^{\text{skin}}(\qskel, \shape)
\right]
(\smpltemplate + \shape \cdot \shapespace + \posedirs(\qskel) ) 
\label{eq:skelskin}
\end{equation}
where $G_i^{\text{skin}}(\qskel, \shape)$ is a rigid transformation that will be defined in Eq.~\eqref{eq:skin_pose}. It translates and rotates the vertices associated with the i-th limb depending on the pose parameter $\qskel$.
 $W_{i}^{\text{skin}}$ is a $6890 \times \osimNbparts$ matrix \changed{of skinning weights} indicating how the vertices of the \smpl mesh are affected by each rigid transformation $i$. \changed{Those weights are inherited from SMPL, by defining a corresponding SMPL joint for each of the $\osimNj=\osimNbparts$ joints of \skel.}

To define the transformations $G_i^{\text{skin}}$ we use the composition of rigid transformations $T(\mat{R}, \vect{t})$ defined by a rotation matrix $\mat{R}$ and a translation $\vect{t}$, 
as well as \changed{per-joint local} transformations $G^B_{k}(\qskel_i, \shape)$, which are pure rotations for most joints, and a combination of rotation and translation for the spine and shoulder blades as explained in \secref{sec:osim}.
The \changed{global transformation to apply to the skin vertices} is computed as $G_i^{\text{skin}}(\qskel, \shape) =$ 
\begin{equation}
\begin{aligned}
\changed{\prod_{k=0}^{i}} \;
\loctosmplcol{T(R_k(\shape), \jskel_k(\shape))}
\; G^B_{k}(\qskel_k, \shape) 
\; \smpltoloccol{T(R_k(\shape), 0)^{-1} \; T(0, \jskel_k(\shape))^{-1}}
\end{aligned}
\label{eq:skin_pose}
\end{equation}
The green term transforms the i-th limb vertices back to the unposed bone space, by centering it on its joint location \smpltoloccol{$T(0, \jskel_k(\shape))^{-1}$}, and undoing the T-pose bone rotation \smpltoloccol{$T(R_k(\shape), 0)^{-1}$}. Then, the joint-specific transformation $G^B_{k}(\qskel_k, \shape)$ is applied. Finally, the bone vertices are posed back to SMPL's posed space by applying the rotation \loctosmplcol{$R_k(\shape)$} and the translation \loctosmplcol{$\jskel_k(\shape)$}. $\jskel_k(\shape)$ is the k-th joint location in T-pose ($\qskel=0$) as defined in \changed{Eq.}~\eqref{eq:skeljoints}. The leading product enforces the kinematic tree structure.

\changed{The pose-dependent deformations of SKEL are inherited from \smpl. For each degree of freedom of \skel, we define a corresponding degree of freedom of \smpl and transfer the pose-dependent deformations ${\posedirs}_{i}(\qskel_i)$. For \skel's joints that do not have an equivalent joint in \smpl, we default to linear blend skinning with no pose correctives.} 
While this transfer is not optimal and creates artifacts in extreme poses (see Sup.~Video), \skel can 
match \smpl meshes with an average vertex-to-vertex error below 3 cm; see \figref{fig:fit_err_skel}. 
We leave the learning of \skel-specific pose-dependent deformations using \db for future work.

{\bf Joints.} \changed{\skel's unposed joints are regressed from the unposed skin vertices $\skelskinverts(\shape, \qskel=0)$ with the learned anatomical joint regressor $\jointreg$, to get the unposed joints $\josim(\shape)$.} Those joints are then posed with the parameter $\qskel$, like the skin vertices, by applying the rigid transformations $G_i^{skin}$: 
\begin{equation}
\jskel(\shape, \qskel)
=
\left[ \changed{\sum_{i=1}^{\osimNj}} W_{i}^J G_i^{skin} (\qskel, \shape) \right]
\josim(\shape)
\label{eq:skeljoints}
\end{equation}
only with different weights $W_{i}^J$ that ensure that the proper joint is affected by the transformation.
Note that for \skel we use a simplified hinge joint at the knee.

{\bf Skeleton.} To obtain the shaped and posed skeleton mesh, a similar equation is used.
We name the initial skeleton template mesh $\osimtemplate$ in which every bone mesh is axis-aligned and \changed{has its parent joint at the world's origin (\figref{fig:osim_markers} right shows the unposed template femur)}.
This mesh is scaled using $s(\josim(\shape))$, a per-bone scaling factor defined by the regressed joint locations, namely the limb lengths they define.
Then the scaled vertices are posed to obtain the posed skeleton vertices 
\begin{equation}
\skelskelverts(\qskel, \shape)
= 
\left[
\changed{\sum_{i=1}^{\osimNj}} W_{i}^{\text{skel}} G_i^{\text{skel}}(\qskel, \shape) \right]
(s(\josim(\shape)) \cdot \osimtemplate)
\label{eq:skelskel}
\end{equation}
where $W_{i}^{\text{skel}}$ are boolean per-bone weights, except for the spine and rib cage where the weights are interpolated \changed{to be 0 at the bottom of the spine section and 1 at the top}.
The skeleton vertex transformations are computed with 
\begin{equation}
\begin{aligned}
G_i^{\text{skel}}(\qskel, \shape) = \changed{\prod_{k=0}^{i}}\: & 
\loctosmplcol{ T(0, \jskel_k(\shape))\:
T(R_k(\shape), 0) }\:
G^B_{k}(\qskel_k, \shape)
\end{aligned}
\label{eq:skeleton_pose}
\end{equation}
in which the unposed bone mesh is transformed by the joint transformation $G^B_{k}(\qskel_k, \shape)$, then oriented with \loctosmplcol{$R_k(\shape)$} to be aligned with the limb's skin and translated to its T-pose joint $\loctosmplcol{\jskel_k(\shape)}$.

\begin{figure}[t]
    \centering
    \begin{subfigure}[]{0.9\columnwidth}
        \centering
        \includegraphics[height=3cm]{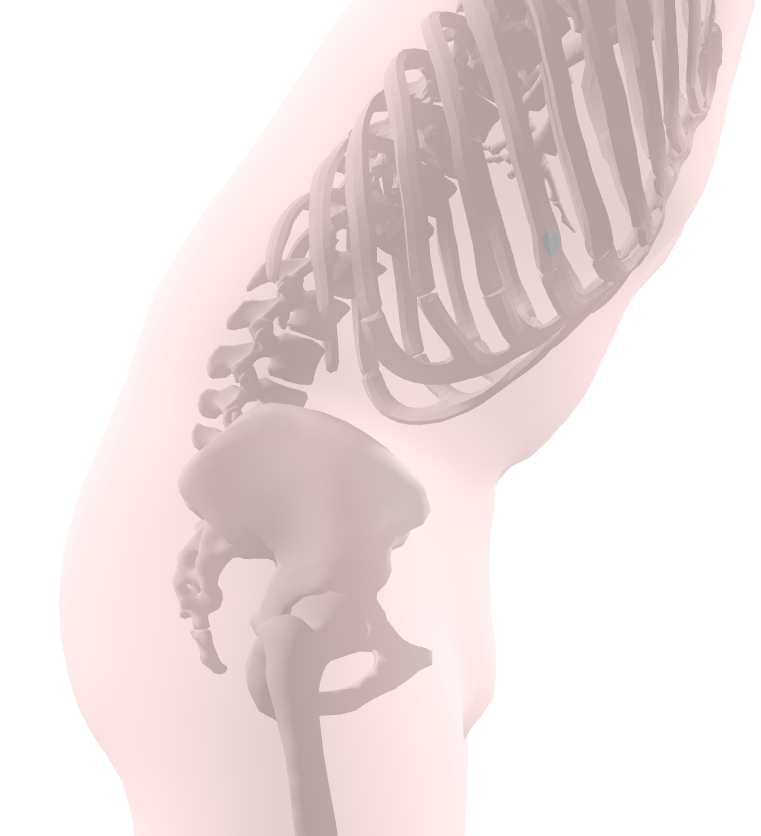}
        \includegraphics[height=3cm]{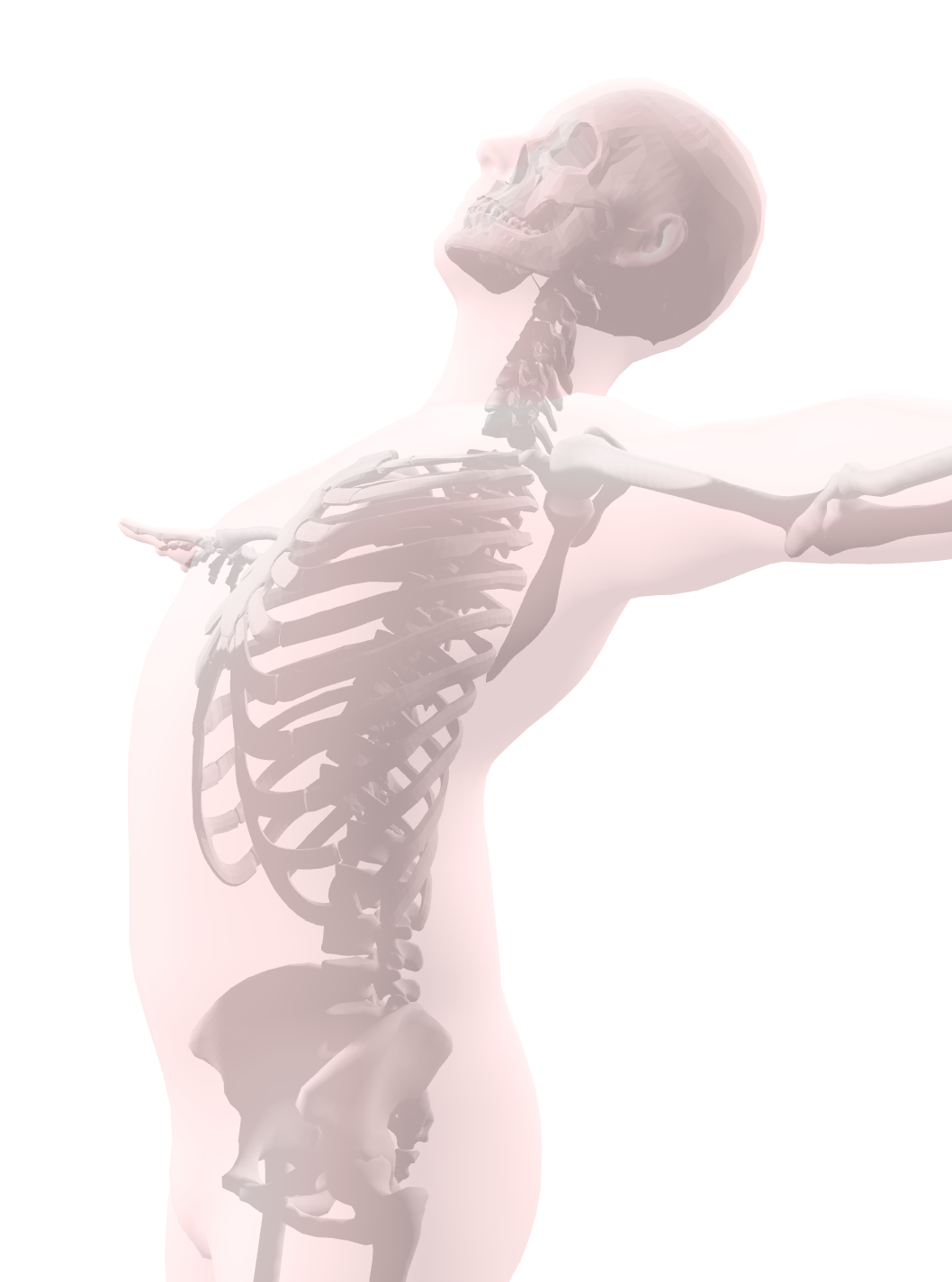}
        \includegraphics[height=3cm]{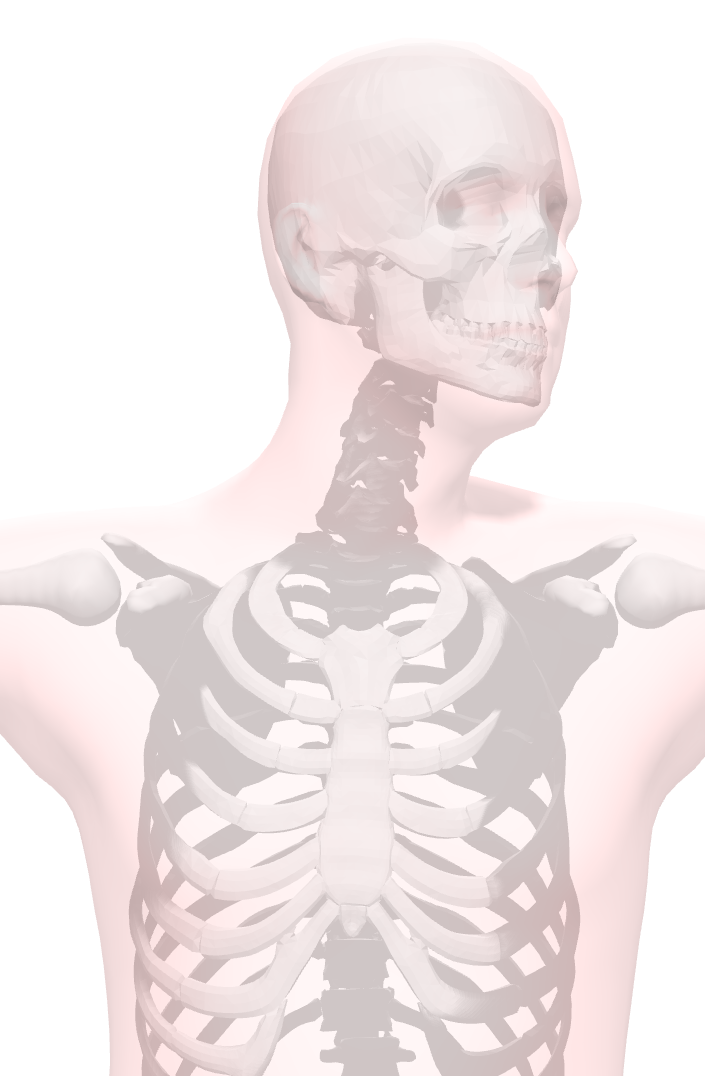}
        \caption{Lumbar flexion, thorax extension, and head twist.}
        \label{fig:spine}
    \end{subfigure}
    \begin{subfigure}[]{0.9\columnwidth}
        \centering
        \def\w{0.32\columnwidth}
        \includegraphics[width=\linewidth]{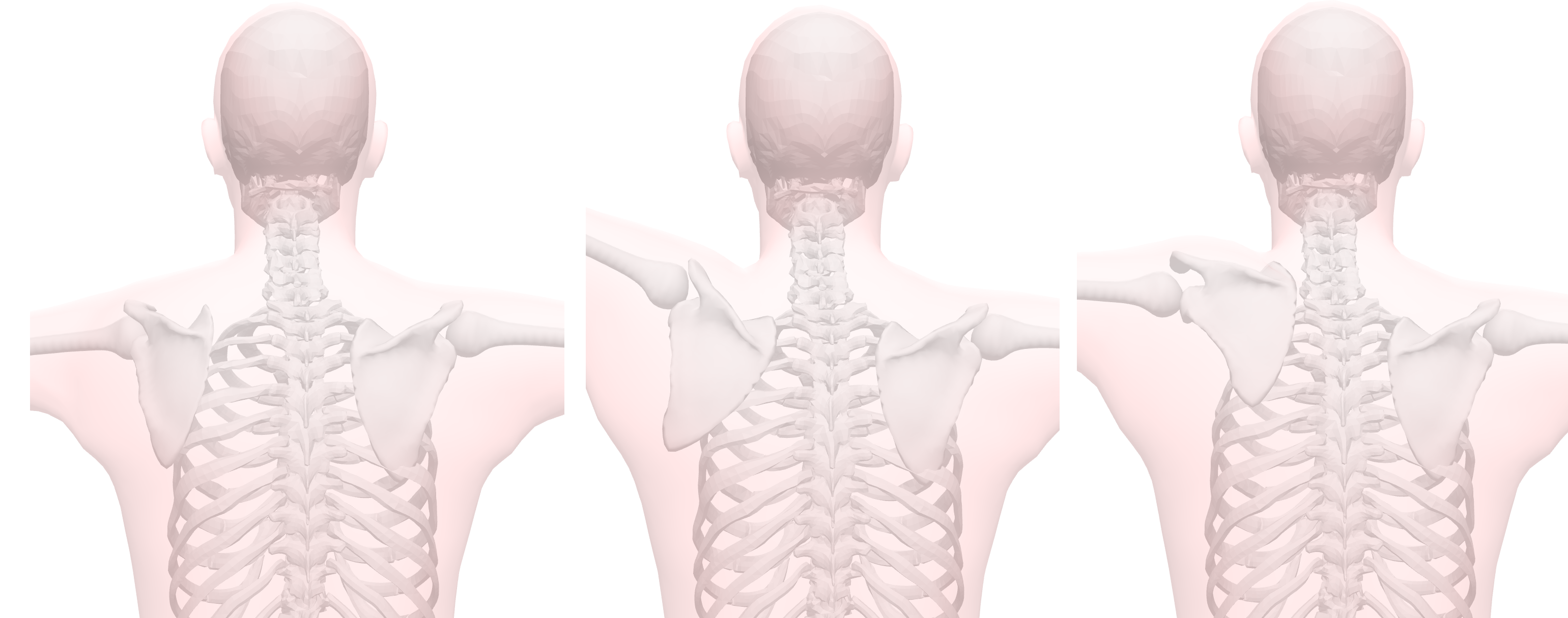}
        \caption{Scapula's abduction, elevation, and upward rotation.}
        \label{fig:scapula}
    \end{subfigure}
    \begin{subfigure}[]{0.9\columnwidth}
        \centering
        \includegraphics[width=1\columnwidth]{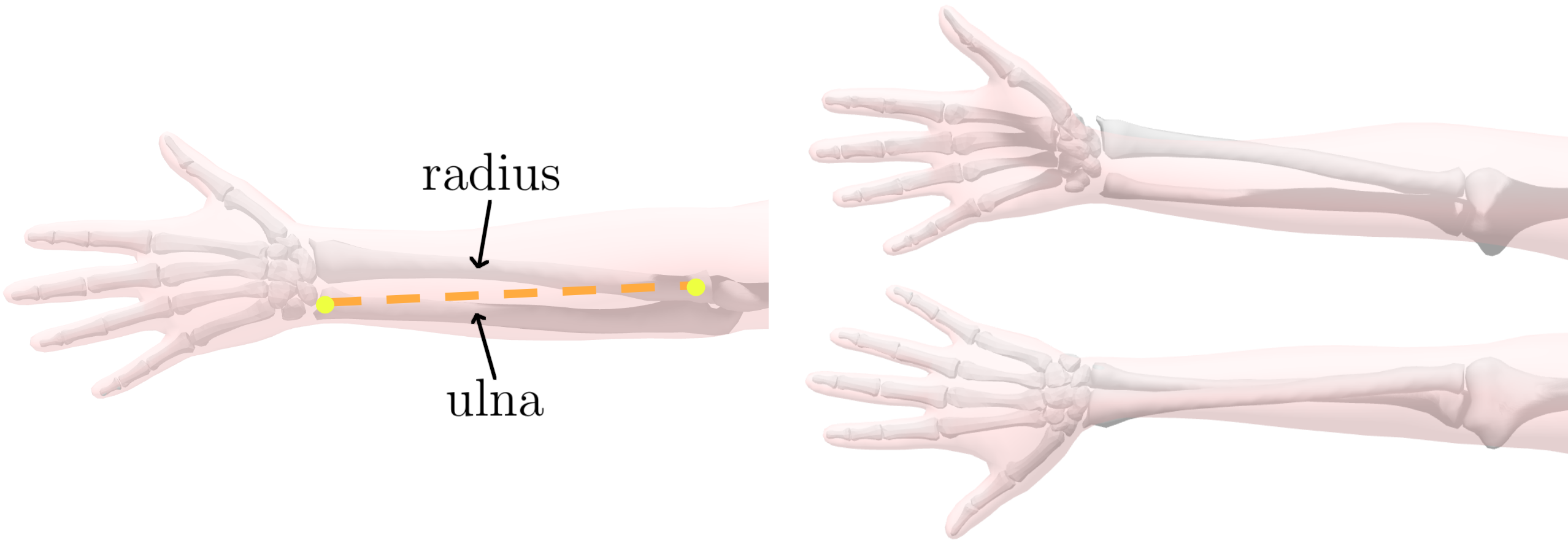}
        \caption{Left: axis of rotation of the radius (lateral view). Right: forearm supination and pronation (top view).}
        \label{fig:forearm}
    \end{subfigure}  
    \caption{Illustration of \skel's degrees of freedom. The bone and body surface meshes are controlled by the same kinematic tree.}
    \label{fig:skel_dof}
\end{figure}

\changed{Finally, we define the range of possible angles for specific degrees of freedom like the shoulder blades, knee, arms, and spine motions}.
Figure \ref{fig:skel_dof} illustrates SKEL's degrees of freedom for the spine, shoulder blades, and arm pronation.
\changed{Note that the deformation of the body surface (pink) is driven by the \osim pose, thus combining the \smpl surface model with an anatomical skeleton.}

\section{Evaluation}

In this section, we evaluate the fit accuracy of the \db dataset,
 the learned anatomical joint regressors, and the skeleton meshes obtained by fitting \skel to \smpl meshes.

\begin{figure}[tbh]
      \centering
      \includegraphics[width=1\columnwidth]{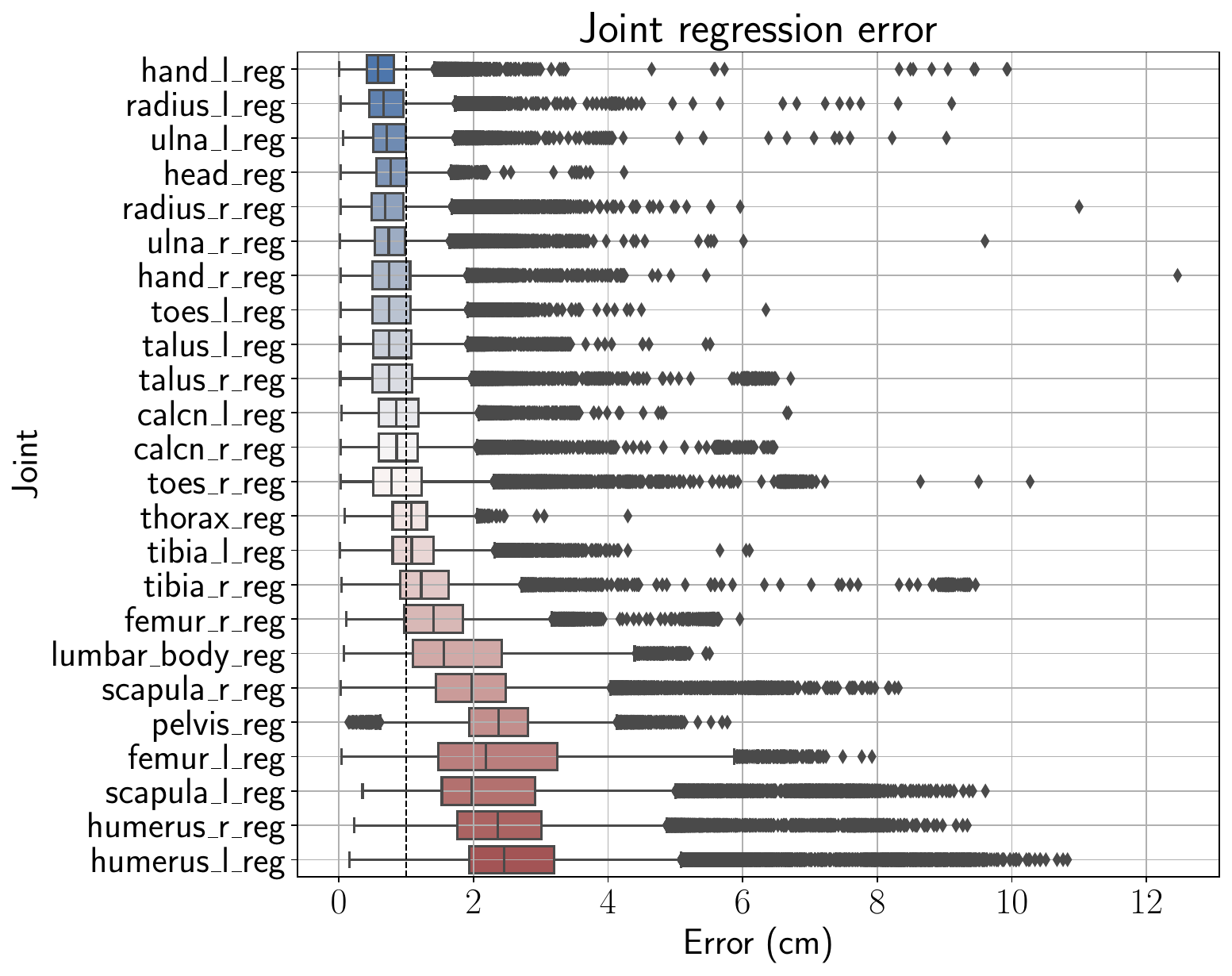}
      \caption{\changed{Anatomical joints regression error over the female DFAUST dataset.}}
      \label{fig:jointposerr_all}
\end{figure}

\subsection{Evaluating \db fits}
\label{sec:dataset_eval}

In \secref{sec:dataset} we simulate optical motion capture markers on \smpl sequences and fit the \osim biomechanical skeleton to them. 
We evaluate these fits by computing the Mean Absolute Error (MAE) between the target and the fitted markers. In \tabref{tab:markers}, for each subset of AMASS, we report the average error of bony and soft markers across all frames.
For comparison, these distances are similar to the body shape reconstruction error from markers reported in \cite{Loper:SIGASIA:2014} and significantly more accurate than the held-out marker error \cite{Loper:SIGASIA:2014}.

\begin{table}[b]
\caption{Marker fitting error of the \osim model on the AMASS dataset.}
\label{tab:markers}
\centering
\begin{tabular}{lcc}
\toprule
MAE in (cm)       & Bony markers & Soft markers \\
\midrule
DFAUST            & 1.54              & 2.00              \\
CMU               & 1.70              & 2.37              \\
MPI\_Limits       & 1.70              & 2.37              \\
\bottomrule
\end{tabular}     
\end{table}

\subsection{Joint regressors}

We evaluate the regressors learned in \secref{sec:learning} on unseen body meshes by comparing the regressed values with the reference \osim alignment.
We train our anatomical joint regressors on the CMU \cite{CMU_mocap_db} and MPI\_Limits \cite{Akhter:CVPR:2015} datasets, which are part of AMASS \cite{mahmood2019amass}. CMU contains good variation in body shape, while MPI\_Limits contains extreme poses. Once trained, we evaluate our regressor on the DFAUST dataset \cite{bogo2017dfaust}, with various motion sequences for 10 subjects with diverse BMIs; DFAUST contains precise SMPL fits to 3D scan sequences.

For each frame of the DFAUST dataset, \db provides the anatomical joint locations $\josim$ that we consider ground truth.
\changed{Then, from the frame's SMPL mesh, we use our learned joint regressor to regress the anatomical joint location $\jregressed$.
In \figref{fig:jointposerr_all} we report the per joint regression errors $|\jregressed_i -\josim_i$|, which are below a centimeter for most joints.
Some joints, such as the humerus, have higher errors.
We inspected the outlying frames and observed some failure cases of the \ab fits for the shoulder joints, which explains the higher values.} The regressed anatomical joints are, in these cases, more plausible than those obtained with \osim, as we show in the supplementary video.

\begin{figure}[tbh]
      \centering
      \includegraphics[trim={0 0 0 0}, clip, width=0.49\columnwidth]{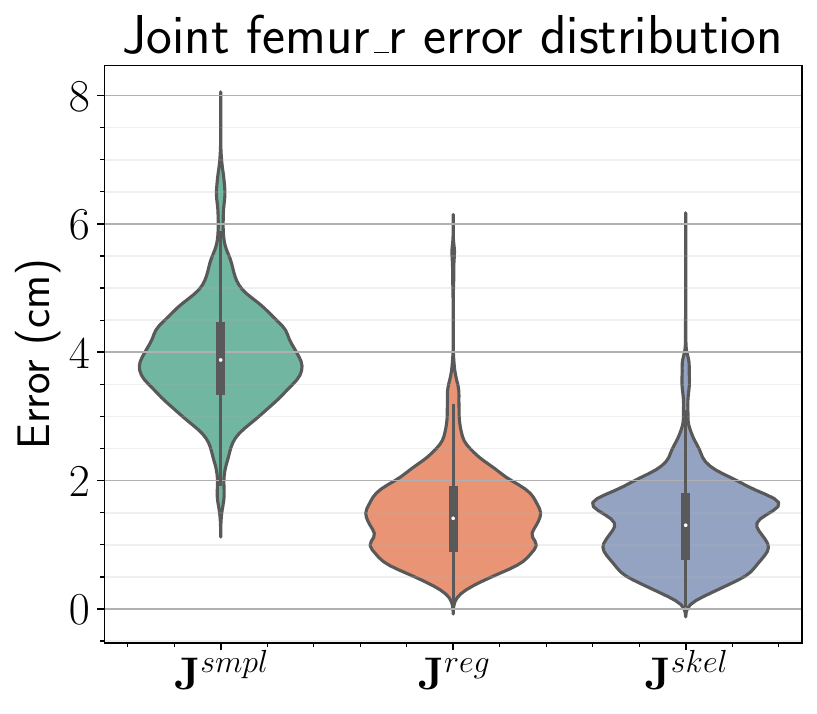}
      \includegraphics[trim={0 0 0 0}, clip, width=0.49\columnwidth]{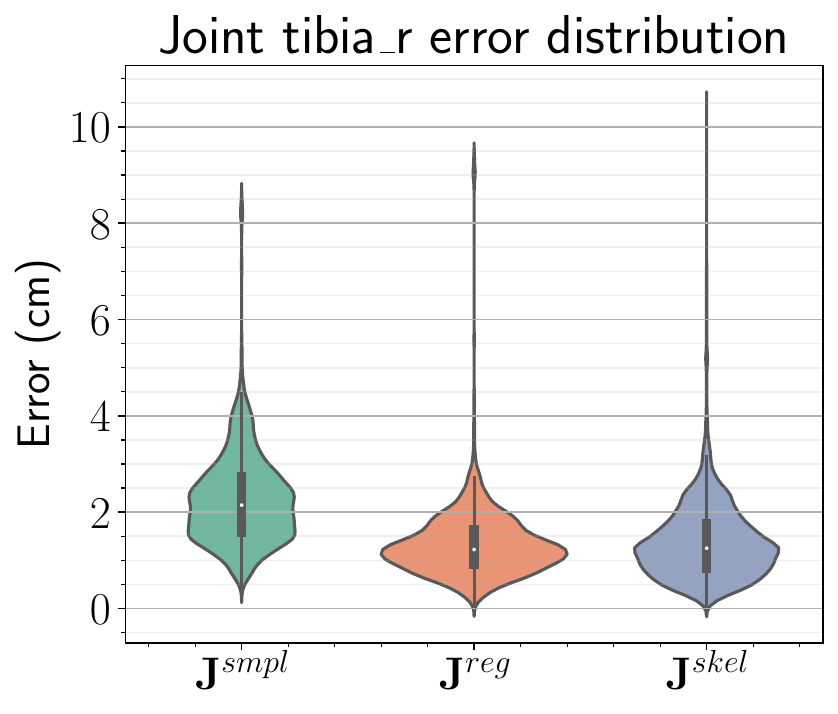}
      \caption{\changed{
      On DFAUST female subjects, we predict the joint locations and show the Euclidean distance errors wrt the ``ground truth'' \osim joint location for the right femur (left) and right tibia (right). We compare 3 methods: $\jsmpl$: using the joints directly from the SMPL fit to the DFAUST bodies. $\jregressed$: joint regressed from the SMPL mesh using our learned anatomical joint regressor. $\jskelfit$: anatomical joints obtained by fitting SKEL to the SMPL mesh.}}
      
      \label{fig:jointposerr}
\end{figure}

\changed{Further, we evaluate the femur and tibia joint location given by different methods as shown in \figref{fig:jointposerr}. 
We consider $\josim$ as the ground truth joint locations and compute the 3D Euclidean distance error of the joints given by SMPL, $\jsmpl$, the anatomical joints we regress from SMPL, $\jregressed$, and the anatomical joints, $\jskelfit$, obtained by fitting fit \skel to the SMPL mesh.
} 
As expected, the \smpl joints have higher error compared to \changed{the learned anatomical ones}.

\subsection{\skel fits to \smpl}

Since \skel has the same \changed{surface mesh topology and shape parameters $\shape$} as \smpl, it can be directly fit to existing \smpl meshes by optimizing its \changed{pose} parameters to minimize the vertex-to-vertex error. 

\begin{figure}[tbh]
      \centering
      \includegraphics[height=4.5cm]{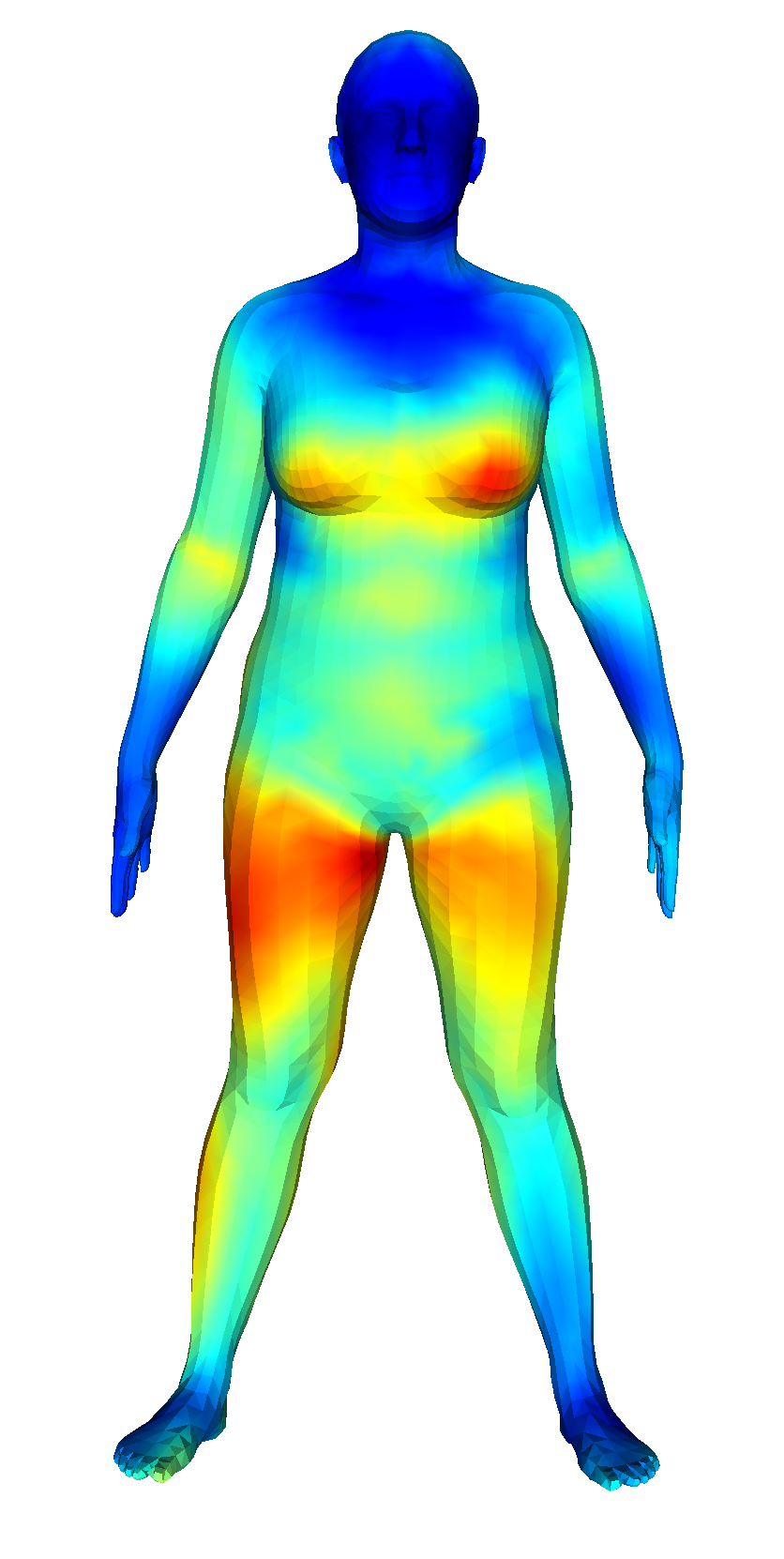}
      \includegraphics[height=4.5cm]{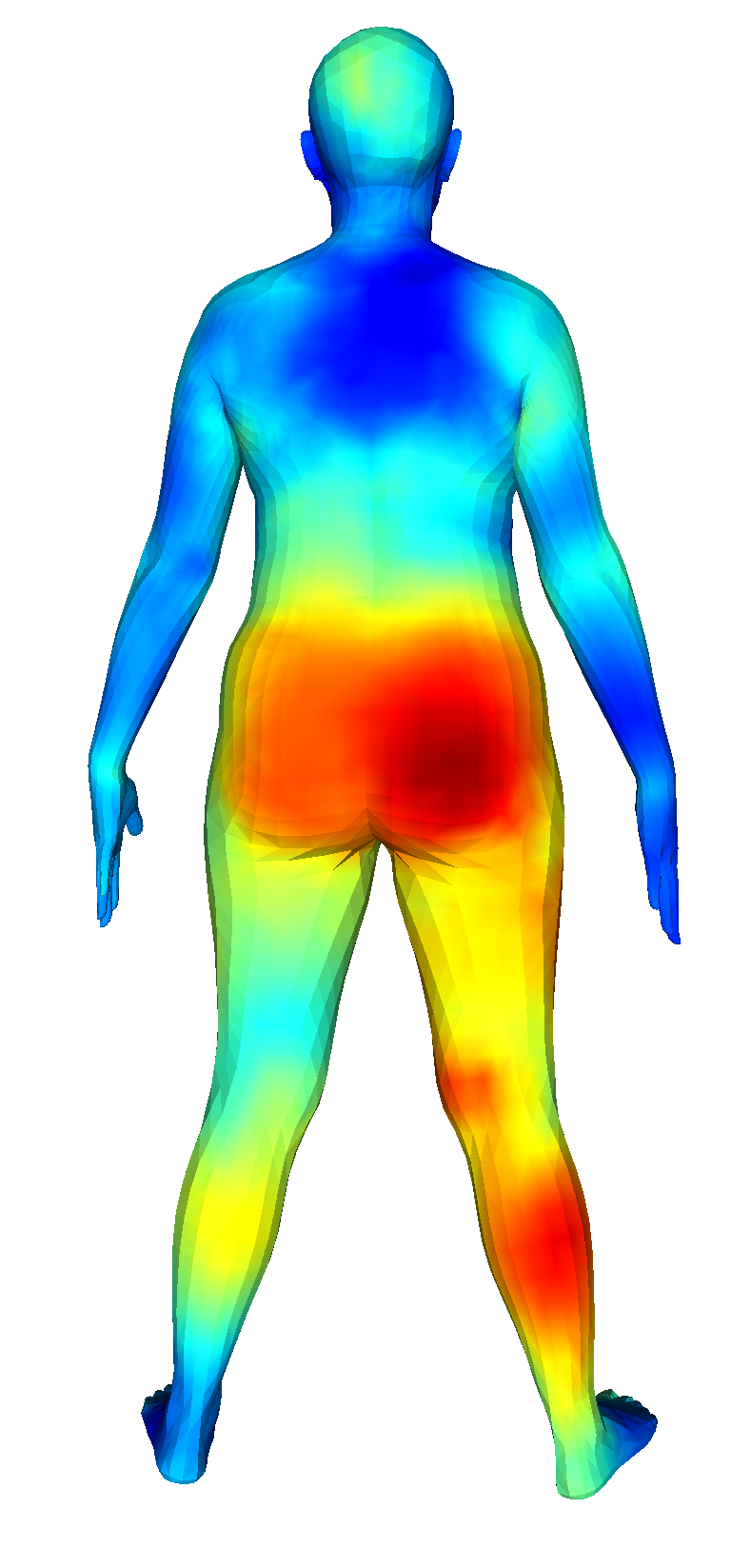}
      \caption{Average per vertex \changed{distance between SKEL and SMPL fit to} the females of the DFAUST dataset.
      Blue: 0 cm, Red 2cm.}
      \label{fig:fit_err_skel}
\end{figure}

To quantitatively evaluate \changed{how similar 
\skel shapes are to \smpl}, \changed{we consider motion sequences from the DFAUST dataset and their \smpl fits with 10 shape parameters}. We fit \skel to each of these \smpl meshes by optimizing its pose parameters $\qskel$.
To evaluate the mesh fits, we compute the mean absolute difference (MAD) between \changed{\skel skin vertices and the target \smpl vertices}, and then average over all the frames.
For males, we find an average difference of 1.1 cm and an average max difference of 2.5 cm, while for females we obtain an average mean difference of 0.9 cm and max of 1.9 cm.
A visualization of these differences on the \smpl body mesh is shown in \figref{fig:fit_err_skel}.
The larger differences can be explained by the approximate pose-dependant blend shapes inherited from \smpl, which could be retrained in future work. 

Fitting \skel to \smpl provides joint locations with similar accuracy as the regressed ones, as reported on \figref{fig:jointposerr}.
Let us note that direct joint regression is faster than estimating the \skel model fit. Applications that require the joint locations but not the skeleton pose parameters, 
and for which time is critical, should prefer the direct regression approach. 

{\it Upgrading \smpl datasets with \skel. }
Since \skel is compatible with \smpl, we \changed{can fit \skel} to SMPL meshes from the 3DPW dataset \cite{von2018recovering} and the synthetic BEDLAM \cite{bedlam} dataset \changed{(\figref{fig:datasets_extention})} \moved{[Figure added]}. The full sequences are shown in the supplementary video.
This effectively upgrades these datasets to include biomechanical pose parameters.

\begin{figure}[tbh]
      \centering
      \def\w{0.32\columnwidth} 
      \def\h{4cm} 
      \includegraphics[height=\h, trim={0 13cm 5cm 13cm}, clip]{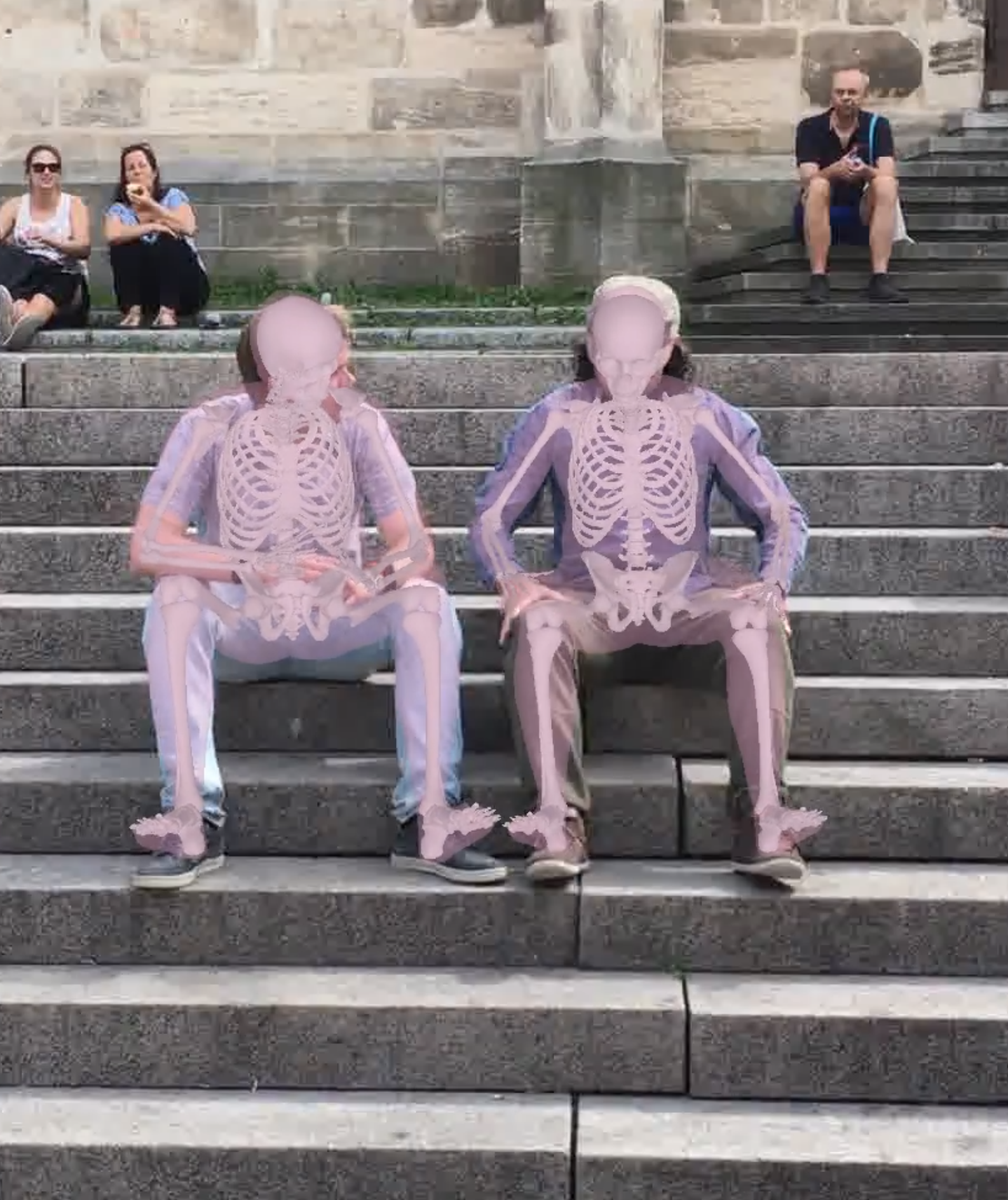}
      \includegraphics[height=\h, trim={6.5cm 10cm 5cm 10cm}, clip]{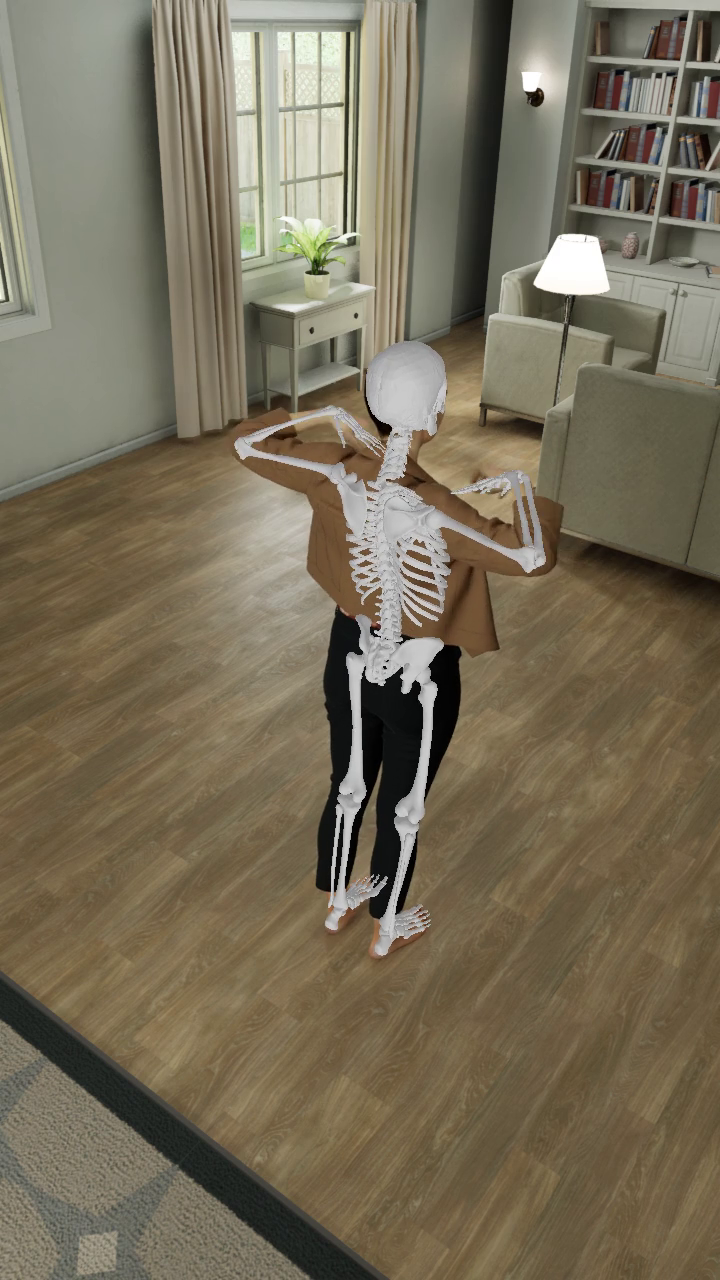}
      \caption{\changed{\skel can be fit to existing SMPL datasets to upgrade them with biomechanical pose parameters. Left: \skel skeleton mesh on a frame of 3DPW \cite{von2018recovering}. Right: \skel skeleton mesh on a frame of BEDLAM \cite{bedlam}.}}
      \label{fig:datasets_extention}
\end{figure}

\subsection{Qualitative comparisons with OSSO}
\begin{figure}[tbh]
      \centering  
      \def\w{0.15\textwidth} 

      \includegraphics[trim={6.3cm 0 6cm 0},clip, width=\columnwidth]{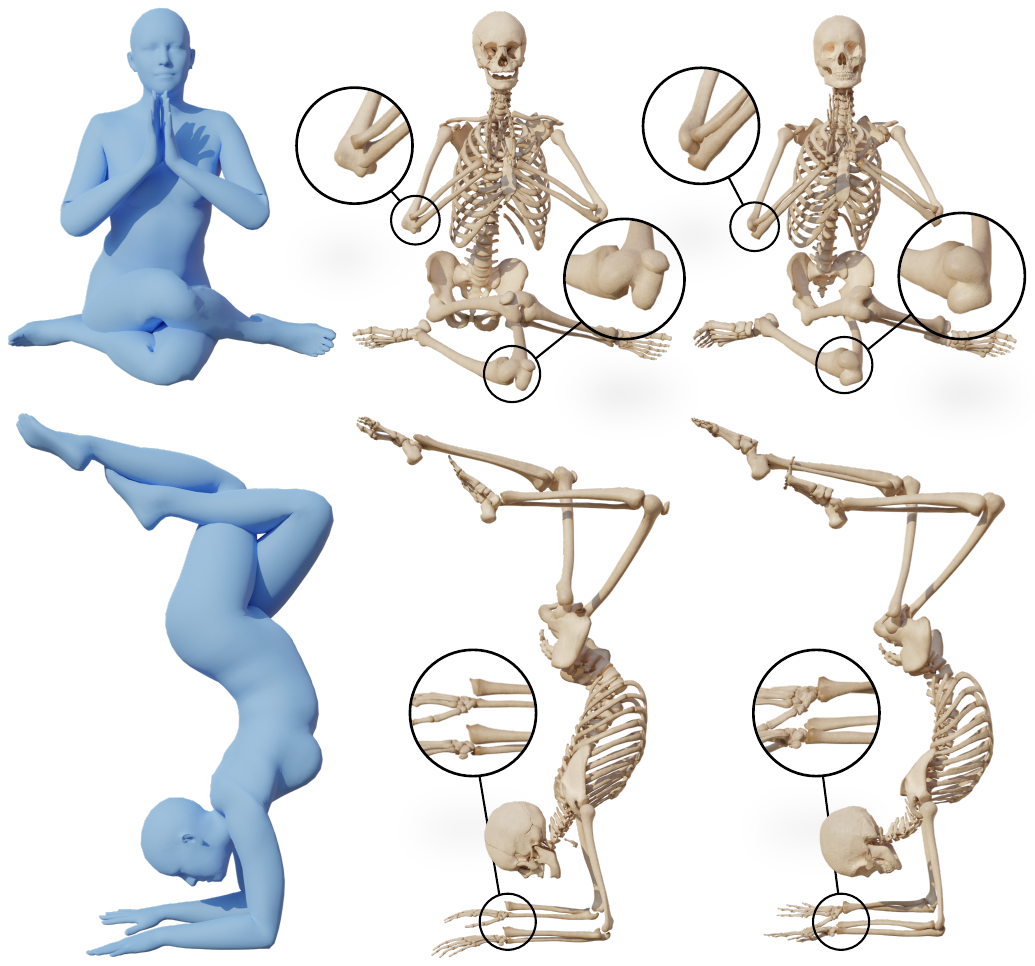}

      \caption{Qualitative comparison between the OSSO and \skel skeletons fitted to MOYO SMPL meshes \cite{moyo}.
      From left to right: Input SMPL mesh, OSSO skeleton, \skel skeleton.
      First row: Due to the anatomic degrees of freedom of \skel, the humerus and femur orientation are properly recovered, while OSSO fails. 
      Second row: OSSO does not model the forearm supination: the radius is not properly rotated with respect to the ulna. The forearm bones have an anatomically correct orientation inside \skel.}
      \label{fig:moyo}
\end{figure}

\moved{[Moved as a separated subsection for clarity]}
\skel fits to SMPL also yield anatomically correct orientations of the bones. To illustrate this we compare the \skel predictions to OSSO skeletons \cite{keller2022osso} on the MOYO dataset \cite{moyo}. 
The \skel skeletons yield more anatomically correct joint location and biomechanically relevant bone angles, as visible in \figref{fig:moyo}; see, for example, the knee orientation as well as arm supination. See Sup.~Mat.~and Sup.~Video for more examples.

\begin{figure}[tbh]
      \centering
      \def\w{0.32\columnwidth} 
      \includegraphics[width=\w]{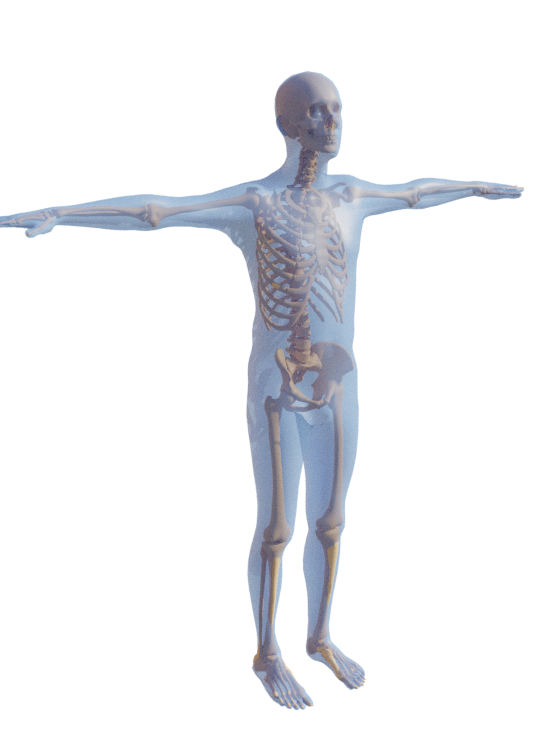}
      \includegraphics[width=\w]{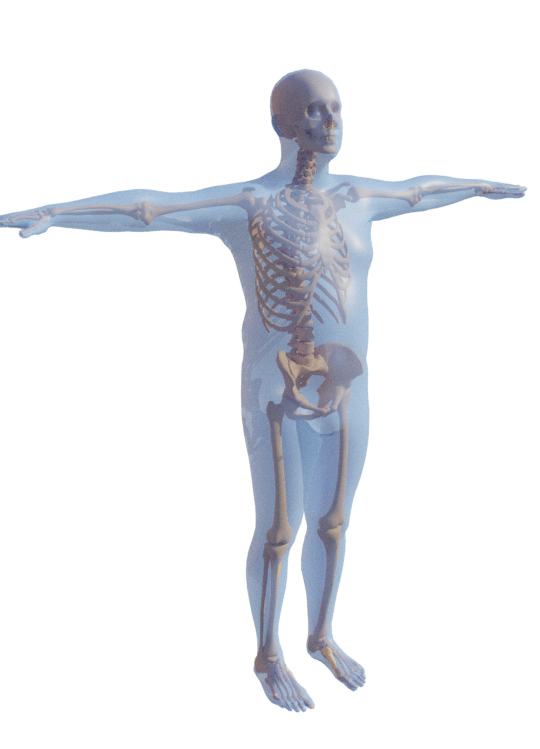}
      \includegraphics[width=\w]{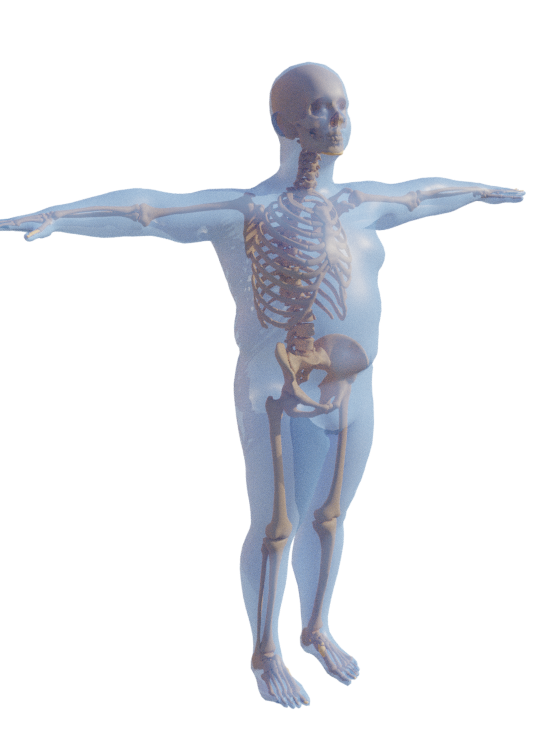}
      \caption{Given an input skeleton, and a target weight, \skel can generate plausible skins while preserving the skeletal structure. From left to right, we set the weight to be 70, 100, and 130 kg.}
      \label{fig:id_variation}
\end{figure}

\subsection{Disentangling body shape and bone lengths}
Since our skeleton mesh is fully defined by the joint segment lengths, we can modify the body shape of a person, while maintaining their skeletal identity. 
This can be helpful for generating a plausible skin mesh from a given biomechanical skeleton. 
\changed{As illustrated in \figref{fig:id_variation}, we optimize the SKEL shape parameters $\shape$ to fit a subject's limb lengths with different target weights. This results in body meshes with different body shapes but the same bone lengths.}

\section{Discussion and Conclusion}

In this paper we describe \skel, a new parametric 3D human body shape model driven by anatomically sound parameters, providing consistent skin and bone geometries.
\skel is learned from \db, a new dataset of skeletons inside SMPL meshes in diverse AMASS poses. We build \db by optimizing \osim, a new biomechanically accurate skeleton model, to fit inside SMPL mesh sequences. Using this paired internal and external data we then learn a regressor from SMPL mesh vertices to the \changed{anatomic} joint locations and bone orientations. 
\skel inherits the shape space from SMPL and the new
anatomic kinematic parameters from \osim. 
From the point of view of vision and graphics, the new model can be used in place of \smpl and it has fewer and more anatomically sound pose parameters (46 for \skel vs 72 for \smpl). This is advantageous, for instance, to regress more accurate anatomic joints from video compared to current approaches solely based on SMPL joints.
From the biomechanics point of view, \skel provides a shape space, which is advantageous to adapt the model to varied body shapes without overstretching certain bones. 
In addition, it provides an animatable model that can take \osim poses and add a \smpl skin for visualization.

\paragraph{\db accuracy limitation.}
\changed{Although the skeletal structures and joint locations computed by \ab are anatomically plausible, they should not be considered as {\em actual} ground truth, but rather a pseudo-ground truth. 
Obtaining actual ground-truth bone measurements of people in motion is not technically feasible.
Thus we rely on marker-based motion capture to obtain estimates of bone motion; this is the current ``gold standard" in biomechanics.
Thus we inherit the accuracy limits of this method, especially for the humerus head prediction, as shown in \figref{fig:jointposerr_all}.
A key next step is to use \skel in the diagnosis of disease and injury and to compare this with traditional motion capture methods.
This is necessary to validate the clinical relevance of the model and methods.
It is worth noting that the learning and rigging pipeline described in \secref{sec:skel} are, in fact, independent of the biomechanical model.
If a new biomechanical model is clinically validated, one can rerun our approach with it to obtain an improved dataset and model.}

\paragraph{\skel extensions and future work.}
There are several directions for extending and improving \skel.
\changed{For instance, in the current model the hands of \skel are rigid. 
Using a biomechanical model with more expressive hands, our approach could be used to put it in correspondence inside SMPL-X \cite{Pavlakos2019smplx}. The bone locations could also be supervised with static medical data observations, such as the ones provided by \cite{wang2019hand}.
}

\changed{Additionally, the current \skel model inherits the 
skinning weights and pose-dependent blend shapes from \smpl. 
These could be retrained from the \db dataset to make the skin surface deformation more accurate.
Ideally, the pose-correctives should be retrained from 3D scan data with \osim as the native parameterization.
This would allow the pose-corrective offsets to be directly learned as a function of \osim parameters.}

Finally, \skel is a step towards a more complete model of the body in motion.
A next step is to add muscle geometry and muscle activation.
For example, we can exploit the estimated \osim skeleton to infer muscle activation using standard biomechanics techniques. This would allow us to upgrade \db with estimated muscle activity.

\paragraph{Conclusion.}
In summary, \skel effectively connects data-driven parametric body shape models with biomechanical skeletons for the first time to enable the integration of these technologies and fields, paving the way towards a new generation of body models and methods that combine the best of both worlds.

\begin{acks}
Michael J. Black (MJB) has received research gift funds from Adobe, Intel,
Nvidia, Meta/Facebook, and Amazon. MJB has financial interests in Amazon,
Datagen Technologies, and Meshcapade GmbH. 
While MJB is a consultant for Meshcapade, his research was performed
solely at MPI.

Soyong Shin performed this work while an intern at the Max Planck Institute for Intelligent Systems.

Marilyn Keller was supported by the International Max Planck Research School for Intelligent Systems. Sergi Pujades’ work was funded by the ANR SEMBA project.

This work uses aitviewer for visualization \cite{kaufmann2022aitviewer}.
\end{acks}

\balance
\bibliographystyle{ACM-Reference-Format}
\bibliography{sample}

\end{document}


\author{Marilyn Keller}
\orcid{0000-0003-2611-8595}
\affiliation{%
 \institution{Max Planck Institute for Intelligent Systems}
 \streetaddress{Max Planck Ring 4}
 \city{T\"{u}bingen}
 \postcode{72076}
 \country{Germany}}
\email{marilyn.keller@tuebingen.mpg.de}
\author{Keenon Werling}
\orcid{0000-0003-3506-7769}
\affiliation{%
 \institution{Stanford University}
 \city{Stanford}
 \state{CA}
 \country{USA}}
 \email{keenon@stanford.edu}

\author{Soyong Shin}
\orcid{0000-0002-0406-7611}
\affiliation{%
 \institution{Carnegie Mellon University}
 \city{Pittsburgh}
 \state{PA}
 \country{USA}}
\email{soyongs@andrew.cmu.edu}
\author{Scott Delp}
\orcid{0000-0002-9643-7551}
\affiliation{%
 \institution{Stanford University}
 \city{Stanford}
 \state{CA}
 \country{USA}}
\email{delp@stanford.edu}

\author{Sergi Pujades}
\orcid{0000-0002-9604-7721}
\affiliation{%
 \institution{Inria centre at the University Grenoble Alpes}
 \city{Grenoble}
 \country{France}}
 \email{spujades@tuebingen.mpg.de}
\author{C.~Karen Liu}
\orcid{0000-0001-5926-0905}
\affiliation{%
 \institution{Stanford University}
 \city{Stanford}
 \state{CA}
 \country{USA}}
\email{karenliu@cs.stanford.edu}
\author{Michael J.~Black}
\orcid{0000-0001-6077-4540}
\affiliation{%
 \institution{Max Planck Institute for Intelligent Systems}
 \city{T\"{u}bingen}
 \country{Germany}}
 \email{mjb@tuebingen.mpg.de}

\renewcommand\shortauthors{Keller, M. et al}

\maketitle

In this document we provide supplementary information regarding the qualitative comparison of \skel to OSSO \cite{keller2022osso}.

\section{Qualitative comparisons with OSSO}
\begin{figure}[b]
      \centering  
      \def\w{0.23\columnwidth} 

      \def\trleC{2} %
      \def\trloC{1.5} %
      \def\trriC{2} %
      \def\trupC{2} %
      
      \def\trleA{2} %
      \def\trloA{11} %
      \def\trriA{6} %
      \def\trupA{2.5} %
      
      \def\trleB{2} %
      \def\trloB{7} %
      \def\trriB{4} %
      \def\trupB{10} %

      \includegraphics[trim={\trleC cm \trloC cm \trriC cm \trupC cm},clip,width=\w]{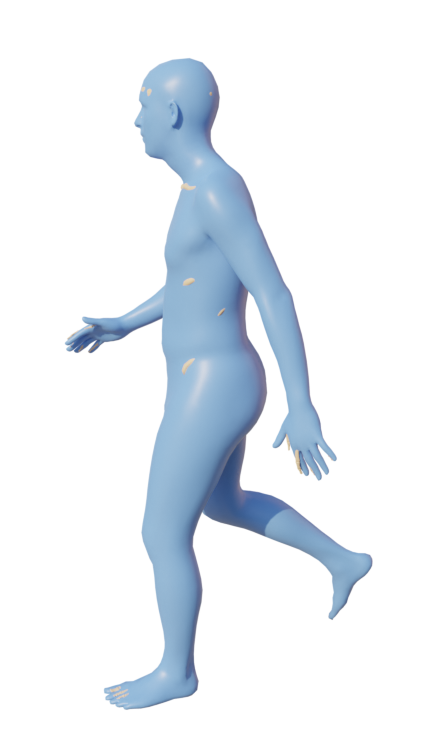}
      \includegraphics[trim={\trleC cm \trloC cm \trriC cm \trupC cm},clip,width=\w]{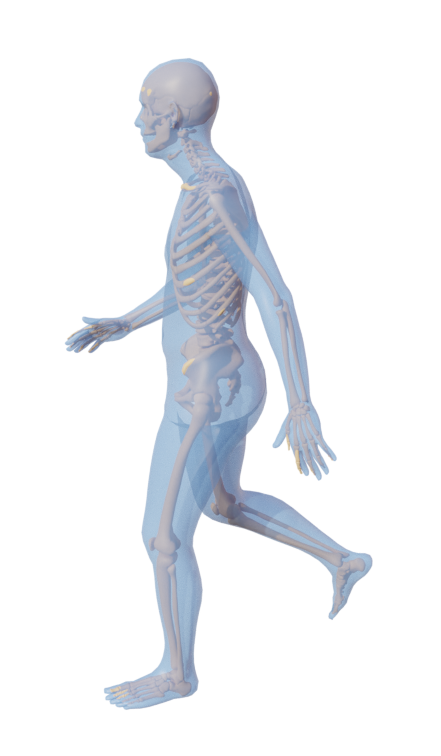}
      \includegraphics[trim={\trleC cm \trloC cm \trriC cm \trupC cm},clip,width=\w]{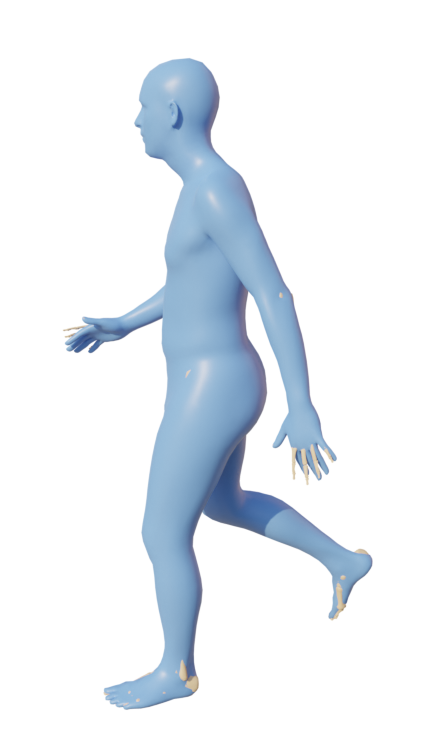}
      \includegraphics[trim={\trleC cm \trloC cm \trriC cm \trupC cm},clip,width=\w]{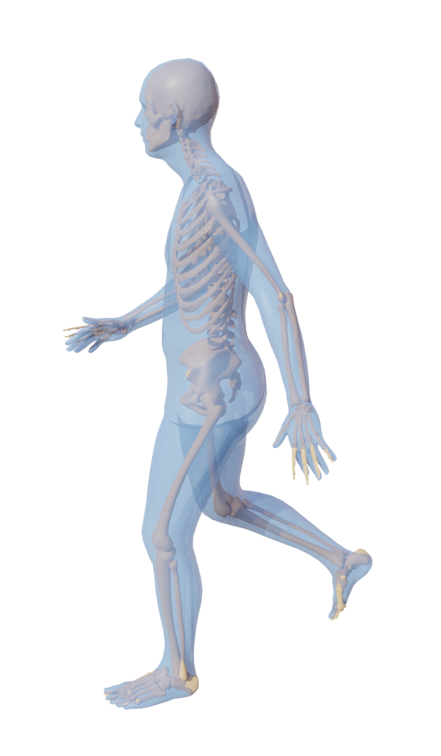}
      
      \includegraphics[trim={\trleA cm \trloA cm \trriA cm \trupA cm},clip,width=\w]{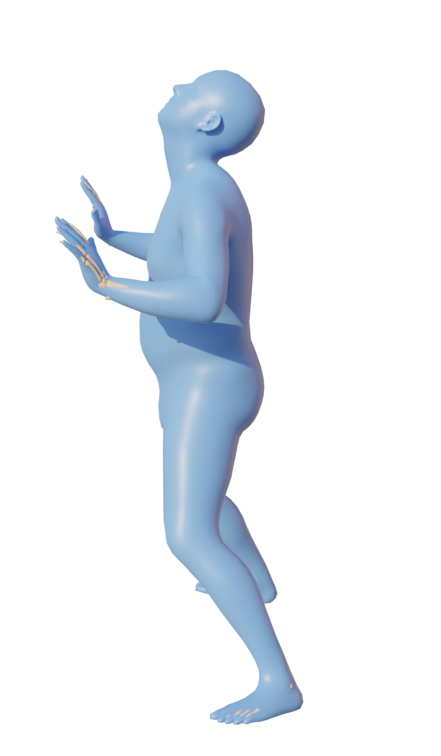}
      \includegraphics[trim={\trleA cm \trloA cm \trriA cm \trupA cm},clip,width=\w]{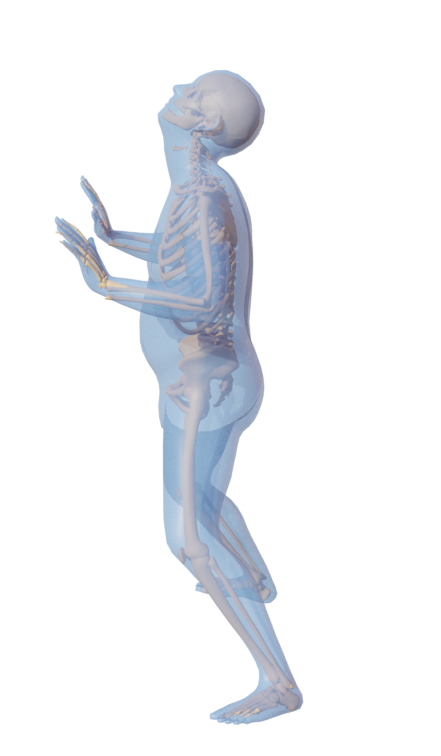}
      \includegraphics[trim={\trleA cm \trloA cm \trriA cm \trupA cm},clip,width=\w]{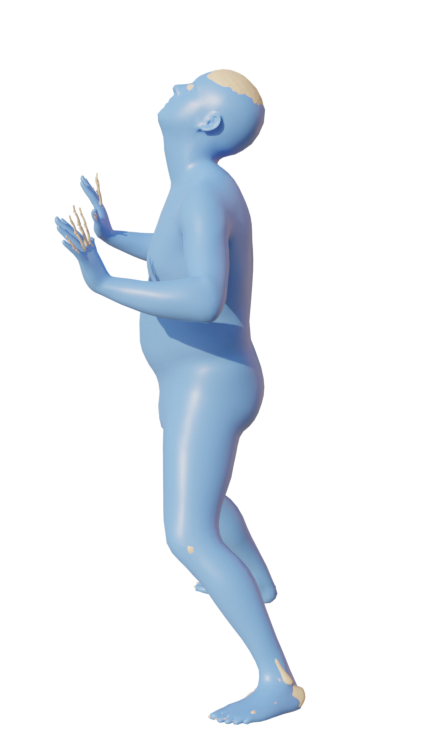}
      \includegraphics[trim={\trleA cm \trloA cm \trriA cm \trupA cm},clip,width=\w]{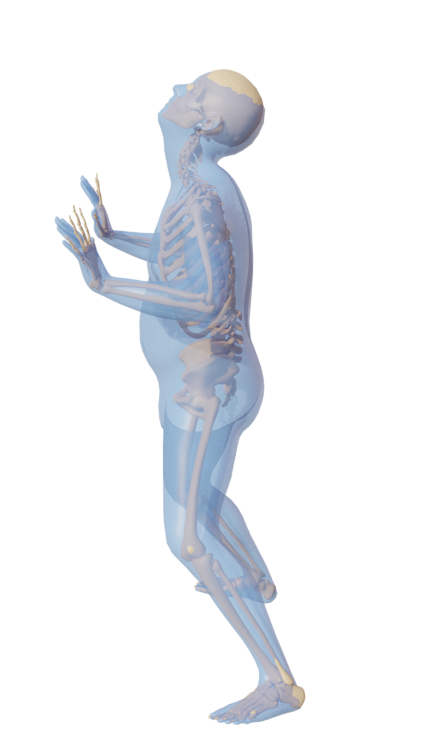}\\      
    \vspace{0.3cm}   
      \includegraphics[trim={\trleB cm \trloB cm \trriB cm \trupB cm},clip,width=\w]{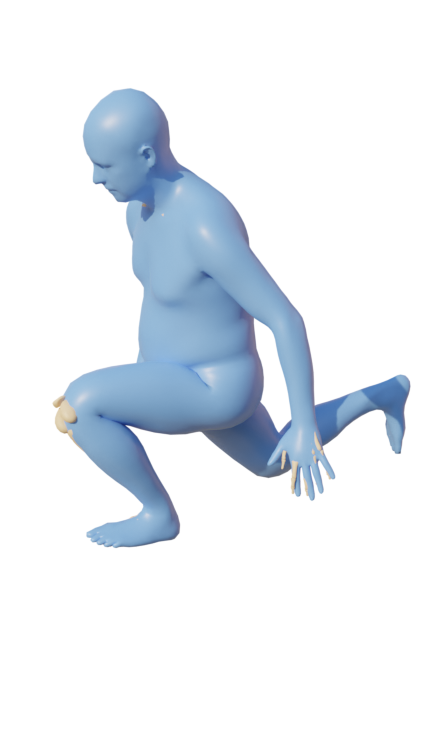}
      \includegraphics[trim={\trleB cm \trloB cm \trriB cm \trupB cm},clip,width=\w]{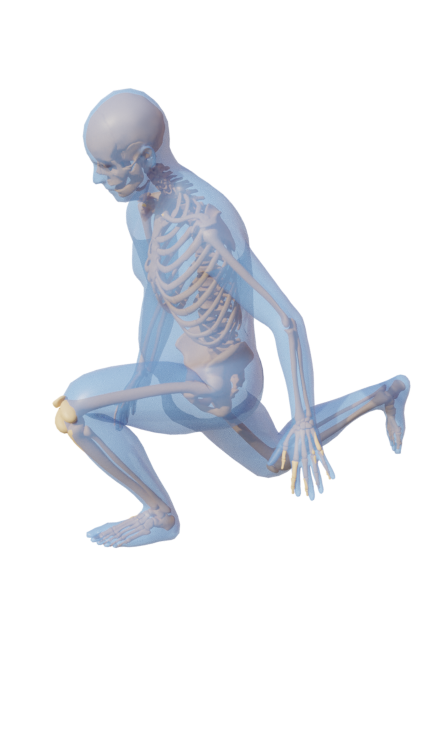}
      \includegraphics[trim={\trleB cm \trloB cm \trriB cm \trupB cm},clip,width=\w]{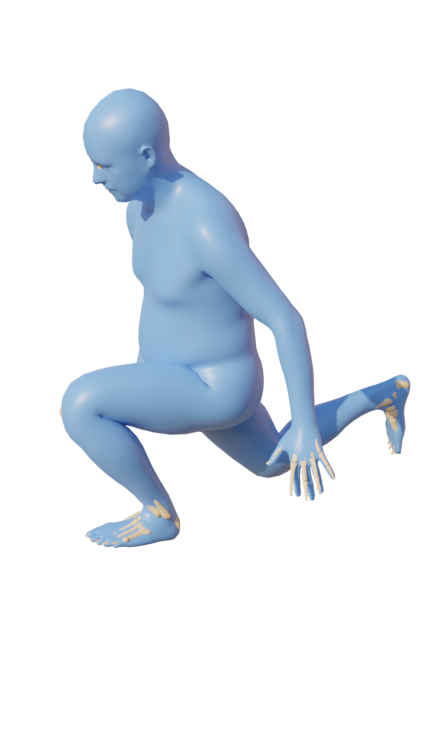}
      \includegraphics[trim={\trleB cm \trloB cm \trriB cm \trupB cm},clip,width=\w]{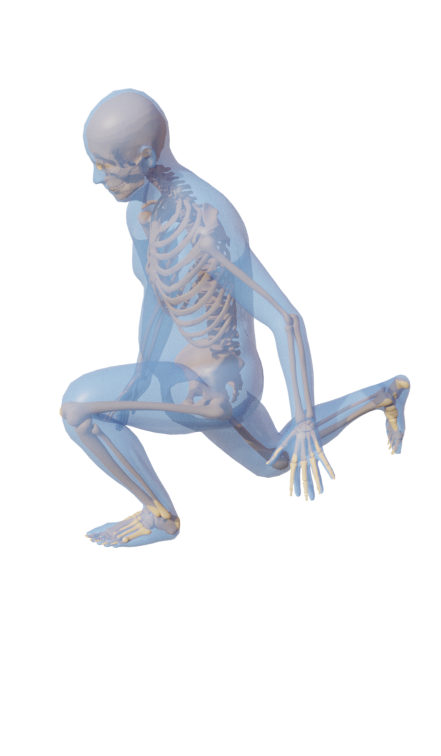}
   
      \caption{Given \smpl meshes from the Total Capture dataset \cite{totalcapture} (in blue) we obtain OSSO \cite{keller2022osso} (left two columns) and the aligned \skel skeleton (right two columns). \skel provides a more anatomically correct skeleton, particularly at the joint level bone orientation, such as the knee and elbow.}
      \label{fig:totalcapture}
\end{figure}

To complement Sec 6.4 of the main manuscript, we provide more qualitative comparisons of the skeletons computed by OSSO \cite{keller2022osso} and fitting \skel to \smpl meshes of the Total Capture dataset \cite{totalcapture}.
The results in \figref{fig:totalcapture} illustrate that \skel provides better bone locations and orientations. This is particular visible in regions such as the ulna and knee where the bones of the articulation have a coherent orientation.

\bibliographystyle{plainnat}
\bibliography{sample}